# Beyond Benefit or Risk: Perceived Impact Profiles of Human-AI Affective Interaction and Their Associations with Psychological Functioning

Lu Chen[a,†1], Fenghua Tang[a,†], Jiayu Zhao[a], Xuanying Li[a], Yanli Wang[a], Weijia Fang[b], Mengyu Miranda Gao[a,*], Zhuo Rachel Han[a,*2]

[a] Beijing Key Laboratory of Applied Experimental Psychology, National Demonstration Center for Experimental Psychology Education, Faculty of Psychology, Beijing Normal University, Beijing 100875, China

[b] Department of Psychology, The Ohio State University, Columbus, OH 43210, USA

## Abstract

Relational AI increasingly serves as an emotional shelter for humans, and its impact is mixed. Prior research has focused on either positive or negative impacts, leaving unclear how they are configured within individuals and relate to psychological functioning. To address these gaps, this study used a sequential mixed-methods design. Study 1 interviewed 52 users with emotional ties to AI and identified four positive impact domains (emotional relief, loneliness alleviation, enhanced interpersonal functioning, and personal growth) and four negative impact domains (virtual-real boundary blur, social replacement, cognitive-emotional reinforcement, and excessive use). Study 2 followed 673 Chinese AI users for six months and identified four profiles of individuals differently impacted by relational AI use: minimal impact, benefit-driven impact, mixed impact, and risk-driven impact. Users in the mixed impact and risk-driven impact profiles were both high in human-AI affective bonding, but those showing risk-driven impact had greater vulnerability, indicated by higher interpersonal need frustration and emotion-regulation difficulties, more depressive and anxiety symptoms, and lower self-esteem and flourishing. Users in the benefit-driven and mixed impact profiles showed more favorable psychological functioning. After controlling for baseline functioning and relevant covariates, Wave 1 profiles did not predict five of the six Wave 2 indicators; only users in the mixed impact profile reported higher flourishing than those in the minimal impact profile. Overall, potential psychological harms associated with relational AI engagement appeared limited and selective. These findings portray relational AI as a heterogeneous socio-emotional context that may partly mirror users' states and traits, warranting individualized, adaptive safeguards.

[1] † These authors contributed equally to this work and share first authorship. E-mail(s): amy_chenlu@mail.bnu.edu.cn; fenghua0604@mail.bnu.edu.cn.

[2] * Co-corresponding authors. E-mail(s): m.gao@bnu.edu.cn; rachhan@bnu.edu.cn.

## 1 Introduction

More than a decade ago, Turkle (2011) argued that people were beginning to expect more from technology and less from one another in their search for connection. Relational AI has made this concern newly concrete. Once largely speculative, relationships with AI are becoming part of everyday social and emotional life. Between 2022 and mid-2025, the number of AI companion applications reportedly increased by 700% (Andoh, 2026), while a market scan of 110 platforms estimated between 1.1 and 2.2 billion monthly visits worldwide (Qian et al., 2025). Affective use is also not limited to dedicated companion applications: large-scale analyses of Claude and ChatGPT conversations have identified emotional support, personal guidance, companionship, and related affective cues in general-purpose systems (Anthropic, 2025; Phang et al., 2025). AI is therefore becoming more than a communication tool; it is beginning to reshape how some users experience companionship, support, and emotional connection.

The appeal of these systems lies partly in their ability to offer responsive and nonjudgmental interaction with little interpersonal friction. They are available on demand, allow users to confide without waiting for another person, and can adapt to preferred relational roles. These features may help users feel heard, accepted, and socially accompanied (Irias et al., 2026). Empirical studies suggest that AI companions can reduce momentary loneliness, particularly when users feel understood by the system (De Freitas et al., 2025). Mental health conversational agents have also shown benefits for depressive symptoms and psychological distress (Li et al., 2023). Relational AI may therefore function as an accessible source of emotional shelter and, in some contexts, as a scaffold for reflection, coping, and social rehearsal.

However, the same form of interaction may also generate risks. Human relationships require reciprocity, disagreement, negotiation, and repair, whereas AI companions can provide persistent affirmation without having needs of their own. Such interaction may subtly alter users' expectations of the comparative value of offline relationships (Andoh, 2026). Longitudinal evidence suggests that turning to social chatbots for companionship may predict greater emotional isolation, although lower social connection can also increase subsequent chatbot use (Folk & Dunn, 2026). Excessive affirmation may further reinforce existing beliefs or emotional states, particularly among vulnerable users (Chandra et al., 2026; Shimgekar et al., 2026). Safety evaluations also indicate that apparently caring responses can sometimes mirror or normalize unsafe content rather than interrupting harmful interaction trajectories (Juneja & Lomidze, 2026).

Taken together, existing evidence suggests that human-AI affective interaction can involve both benefit and risk. However, prior research has largely examined isolated outcomes or average associations, obscuring how benefits and risks are configured within individuals.

Some users may perceive predominantly positive impacts, whereas others may experience strong support alongside salient risks. A second ambiguity is that AI-related impacts and broader psychological functioning are often treated within the same outcome space. Existing psychological functioning may shape how users engage with AI (Folk & Dunn, 2026), while profiles of AI impact may, in turn, carry significance for later psychological functioning. To address these gaps, the present research aims to clarify both the heterogeneity of perceived AI-related impacts and their psychological significance. Specifically, it examines how benefit and risk are configured across users and how these configurations relate to human-AI relationship characteristics and broader psychological functioning over time.

## 2 Literature Review

### 2.1 Human-AI affective interaction and perceived AI-related impacts

Relational AI refers to conversational agents that simulate relational capacities through sustained, responsive, and personalized interaction (Irias et al., 2026). In the present study, the term is defined by the relational function of the interaction rather than by product type, encompassing both dedicated companion applications and general-purpose chatbots when they are used for emotional or relational purposes (Phang et al., 2025). Within this scope, human-AI affective interaction refers to exchanges in which users seek emotional expression, comfort, companionship, or other forms of personally meaningful support. From a relationship-science perspective, relational AI can sustain frequent interactions and respond in ways that users experience as understanding, validating, and supportive (Smith et al., 2025). Consistent with these affordances, relational AI has been found to provide emotional validation and opportunities for self-expression and social rehearsal (Yuan et al., 2025). It can also reduce momentary loneliness, particularly when users feel heard by the system (De Freitas et al., 2025).

However, the same affordances may also create risks. Continuous availability and personalized relational continuity may encourage users to rely on AI for emotional regulation. When the system changes, fails, or becomes unavailable, that reliance can become a source of distress (Laestadius et al., 2024). Highly affirming responses may reinforce users' existing beliefs or emotional states, a process referred to as the "echo-chamber effect" (Starke et al., 2024). For example, sycophantic responses have been found to increase confidence in one's own position and reduce willingness to repair interpersonal conflict, despite being preferred and trusted by users (Cheng et al., 2026). Moreover, AI lacks independent psychological needs and generally requires less negotiation, compromise, and tolerance of disagreement than close human relationships (Smith et al., 2025). Sustained AI companionship may therefore supplement offline human relationships for some users (Xia et al., 2025). At the more severe end, prolonged or sycophantic AI interactions may intensify delusion-related thinking—sometimes termed "AI psychosis," although it is not a recognized diagnosis and current evidence remains simulation-based (Chandra et al., 2026; Shimgekar et al., 2026).

Taken together, prior evidence does not support treating human-AI affective interaction as uniformly beneficial or harmful. In the present study, we use perceived AI-related impacts to describe these benefits and risks that users attribute to their affective interactions with AI. I*n* addition, these perceived AI-related impacts may coexist within individuals or form different combinations across users (Yuan et al., 2025). However, prior research has largely examined these impacts separately, leaving unclear how they are configured within individuals.

### 2.2 Impact profiles: A person-centered perspective

The coexistence of supportive and risky experiences raises a central question: how are different AI-related impacts organized within the same individual? Existing work typically examines emotional support, personal growth, dependence, problematic use, and social displacement as separate outcomes (Fang et al., 2025; Ho et al., 2025). Such analyses estimate average associations but cannot show whether particular benefits and risks co-occur within individuals. Evidence nevertheless suggests that users differ in both the intensity and form of their AI involvement. Prior research has documented emotional validation and social rehearsal alongside dependence and withdrawal, with experiences also varying across relationship stages (Yuan et al., 2025). Liu et al. (2025) identified seven user clusters based on engagement patterns, psychological characteristics, and social resources. Some users reported greater social confidence, whereas others combined intensive or problematic use with isolation. Consumer-AI relationships also differed in users' purposes, self-congruence, and degree of self-AI integration (Alabed et al., 2024). However, these studies did not define groups through configurations of perceived positive and negative AI-related impacts.

A person-centered approach is suited to this problem because it treats the individual, rather than an isolated variable, as the unit in which multiple characteristics are organized. Variable-centered analyses estimate average associations among variables across a sample (Howard & Hoffman, 2018). Person-centered analyses instead identify subgroups of individuals who share similar configurations across several indicators (Bergman & Magnusson, 1997). These approaches are complementary: variable-centered analyses describe relations among constructs, whereas person-centered analyses examine how those constructs combine within people (Marsh et al., 2009). Profiles may differ in their general level across all indicators, but they may also differ in shape, with similar overall levels arising from different balances of positive and negative impacts.

Applied to perceived AI-related impacts, LPA can test whether supportive and risky experiences form recurring configurations across users. This provides information that cannot be recovered from a total impact score or from separate associations involving each outcome. It can distinguish, for example, between users reporting generally little impact, users reporting support with limited risk, and users for whom support and risk are both elevated. At the same time, latent profiles should not be understood as fixed or naturally occurring user types. Profile solutions depend on the indicators, sample, and model specification, and their selection should consider statistical fit, theoretical meaning, subgroup size, and interpretability (Nylund-Gibson & Choi, 2018). The value of LPA therefore lies in identifying model-based configurations of experience that complement, rather than replace, variable-centered evidence.

### 2.3 Impact Profiles and Broader Psychological Functioning

Beyond the impacts that users directly attribute to AI interaction, affective engagement with AI has also been linked to broader psychological functioning across life contexts. Here, broader psychological functioning refers to individuals' general emotional, interpersonal, and psychological adjustment across life contexts, assessed without requiring them to attribute these experiences to AI. Existing findings are mixed. A four-week social-chatbot study reported reductions in loneliness and social anxiety, although its single-group design limited causal inference (Kim et al., 2025). Over 12 months, greater chatbot use predicted higher emotional isolation, while lower social connection also predicted subsequent use (Folk & Dunn, 2026). Outcomes further vary by interaction mode, use intensity, and user context: heavier use has been linked to less favorable outcomes, whereas positive associations with well-being depend on loneliness, offline friendships, initial states, and modality (Fang et al., 2025; Nakagomi et al., 2026; Phang et al., 2025). Thus, AI interaction appears conditionally rather than uniformly related to psychological functioning.

One explanation for these conditional associations is that users' existing psychological functioning may shape how they engage with AI and which impacts they experience. Cross-sectional studies suggest that social anxiety and low self-esteem are associated with problematic AI use through loneliness, rumination, escapism, or immersion, although their temporal order remains unclear (Hu et al., 2023; Yao et al., 2025). Longitudinal evidence further indicates that anxiety and depression may predict later AI dependence through social and escape motives, whereas AI dependence did not predict subsequent symptoms (Huang et al., 2024). Psychological vulnerability does not necessarily imply harm, however, as users with greater loneliness, lower social support, or insecure attachment may also engage more and benefit more from structured AI interventions (Shoshani et al., 2026). Accordingly, concurrent differences between impact profiles may reflect users' prior characteristics, their responses to AI interaction, or both (Boyd & Markowitz, 2026).

At the same time, perceived impact profiles may carry prospective information beyond users' concurrent psychological functioning. Existing longitudinal research has mainly examined chatbot use intensity or mode, changes in use and loneliness, or temporal relations between psychological symptoms and AI dependence (Fang et al., 2025; Folk & Dunn, 2026; Huang et al., 2024). It remains unclear whether configurations of AI-specific benefits and risks are associated with later psychological functioning after baseline differences are considered. A profile-based longitudinal analysis can therefore test whether perceived impact configurations provide incremental information about subsequent functioning.

## 3 Present Study

Taken together, despite growing evidence on the consequences of affective interaction with conversational AI, three limitations remain. First, evidence on the consequences of affective interaction with conversational AI remains fragmented across specific outcomes, and the broader range of impacts experienced by users has rarely been systematically derived from their own accounts. Second, existing studies have predominantly relied on variable-centered analyses of average associations, providing limited insight into whether positive and risk-related impacts coexist within individuals or form distinct patterns across users. Third, the relation between AI-related impacts and broader psychological functioning remains insufficiently understood.

Accordingly, the present research addressed four questions:

RQ1: What positive and negative impacts do users attribute to affective interaction with conversational AI?

RQ2: Could distinct profiles of perceived AI-related impacts be identified among AI users?

RQ3: Do users from different profiles present different levels of psychological functioning?

RQ4: Do users in different Wave 1 impact profiles differ in psychological functioning approximately six months later, after baseline functioning is taken into account?

To address these questions, we adopted a sequential mixed-methods design. Study 1 used semi-structured interviews and inductive thematic analysis to identify users' perceived positive and negative impacts of affective AI interaction. Study 2 then used a two-wave longitudinal survey. Latent profile analysis was applied to identify distinct configurations of perceived AI-related impacts, followed by comparisons in human-AI relationship characteristics and concurrent psychological functioning. Finally, we examined whether Wave 1 profile membership predicted psychological functioning approximately six months later after accounting for baseline functioning and relevant covariates. By combining qualitative, person-centered, and longitudinal evidence, the present research provides a more differentiated account of the psychological significance of affective interaction with conversational AI.

## 4 Study 1: Qualitative Exploration of Perceived AI-Related Impacts

### 4.1 Methods

#### 4.1.1 Participants and Procedure

The participants were recruited from May to June 2025 through online platforms such as Rednote and WeChat. The inclusion criteria were as follows: (a) being aged 18 or older; (b) having self-reported emotional interactions with AI beyond purely instrumental use, such as confiding personal feelings, seeking emotional support, or viewing AI as a friend or partner; and (c) engaging in voluntary participation with signed informed consent. The prospective participants completed a screening questionnaire that included an item that assessed their typical AI use; those who reported using AI exclusively for instrumental purposes were excluded. From the eligible pool, the participants were purposively selected to ensure sufficient variation in sex, relationship status, perceived AI relationship type, and geographic region. The sample size was guided by the principle of data saturation, with interviews continuing until no new themes emerged (Guest et al., 2006).

Following this procedure, semi-structured interviews were conducted with a final sample of 52 participants, aged 18 to 35 years ($M$ = 23.21, $SD$ = 3.99), of whom 45 were female (86.5%). With respect to their perceived relationship with AI, 17 described it as friendship, and 35 described it as a romantic partnership. In terms of their offline relationship status, 37 (71.2%) were single, 9 (17.3%) were in a committed relationship or were married, and the remainder reported other circumstances. The participants interacted with both general-purpose AI assistants (e.g., Doubao, DeepSeek, ChatGPT, and KIMI) and dedicated AI companion applications (e.g., Talkie and Replika), with the most frequently used being Doubao (27.7%), DeepSeek (24.8%), and Talkie (15.8%).

#### 4.1.2 Interview Protocol

Semi-structured interviews were conducted to explore users' perceived impacts of affective interaction with conversational AI. The interview guide covered participants' general AI interaction experiences and perceived changes in emotional experience, self-understanding, real-world interpersonal relationships, everyday functioning, and mental health. Sample questions included: "Has interacting with AI changed your emotional state or the way you deal with negative emotions?", and "Has interacting with AI affected your relationships or communication with people in real life?" And participants were invited to describe both positive and negative influences. Each interview lasted approximately 25–35 minutes, was audio-recorded with participants' informed consent, and was transcribed verbatim. The full interview protocol is provided in Appendix A.

#### 4.1.3 Data Analysis

The interview transcripts were analyzed using inductive thematic analysis with an iteratively developed codebook (Braun & Clarke, 2006), supported by MAXQDA. The analysis focused on participants' accounts of the perceived effects of affective interaction with conversational AI on their emotional states and mental health, self-perceptions, everyday functioning, and offline interpersonal relationships. Rather than applying predefined nodes or subthemes, the coding framework was developed from the interview data through iterative coding, comparison, and team discussion.

The research team first familiarized themselves with the transcripts through repeated reading. Meaningful segments concerning perceived positive or negative impacts of AI interaction were identified and assigned initial descriptive codes. Coded segments referred to coding records extracted from the interview transcripts; when a passage contained meanings relevant to more than one impact, multiple coding was allowed. Through iterative comparison, codes with similar meanings were refined and organized into discrete nodes. The resulting 44 nodes were subsequently clustered into eight subthemes according to their shared conceptual properties. These subthemes were then integrated into two overarching themes: Positive Perceived Impacts and Negative Perceived Impacts.

To assess coding consistency, two members of the research team independently recoded 10 randomly selected transcripts (approximately 19% of the sample) using the established codebook. Agreement rates across the 10 transcripts ranged from 66.67% to 100%. Coding discrepancies were discussed by the research team, who refined the code definitions and reached consensus on the final coding decisions before proceeding to subsequent stages of analysis (Campbell et al., 2013). The finalized coding scheme was then used to review and confirm coding across the full dataset.

Frequencies were calculated at both the coded-segment and participant levels. Participant counts were based on unique participant IDs, such that a participant was counted only once within each node, subtheme, or overarching theme regardless of the number of relevant segments. Participant percentages used the full interview sample as the denominator ($N = 52$). Illustrative quotations were translated from Chinese into English and checked against the original transcripts to preserve their intended meaning.

### 4.2 Results

The analysis yielded 252 coded segments, which were organized into 44 nodes and further grouped into eight subthemes. Four subthemes reflected positive perceived impacts: emotional relief, loneliness alleviation, enhanced interpersonal functioning, and personal growth. The remaining four reflected negative perceived impacts: virtual-real boundary blur, social replacement, cognitive-emotional reinforcement, and excessive use. Together, these subthemes formed two overarching themes: Positive Perceived Impacts and Negative Perceived Impacts. Detailed coding results are presented in Appendix B.

Positive Perceived Impacts accounted for 198 coded segments (78.6%) and were reported by 49 participants (94.2%), whereas Negative Perceived Impacts accounted for 54 coded segments (21.4%) and were reported by 23 participants (44.2%). All participant percentages were calculated using the full interview sample as the denominator ($N$ = 52). Because participants could report more than one type of impact, participant percentages were not mutually exclusive. Notably, 22 participants (42.3%) reported both positive and negative impacts, suggesting that perceived benefits and risks often coexisted within users' experiences.

### 4.2.1 Positive Perceived Impacts

***Emotional Relief.*** Emotional relief was one of the most frequently reported benefits, with 35 participants (67.3%) describing affective interactions with AI as helping them regulate or alleviate negative emotions (63 coded segments, 25.0% of all coded segments; Nodes 1–5). Participants described feeling emotionally understood or "seen" (8 segments; $n$ = 6, 11.5%) and receiving comfort, reassurance, or an outlet for emotional release after disclosure (21 segments; $n$ = 18, 34.6%). Others reported feeling happier or more positive after interacting with AI (23 segments; $n$ = 19, 36.5%) or regaining a sense of calm (7 segments; $n$ = 7, 13.5%). A smaller number described feeling loved or emotionally uplifted through romantic interaction (4 segments; $n$ = 2, 3.8%). As one participant said, AI served as an always-available "emotional tree hole": "Whenever I want to talk, whether late at night or at any time, it can help guide my emotions. I feel that it makes my emotions more stable" (A26).

***Loneliness Alleviation.*** Loneliness alleviation was reported by 11 participants (21.2%) and comprised 17 coded segments (6.7% of all coded segments; Nodes 6–7). Participants described AI as reducing feelings of loneliness, emptiness, or lack of companionship. Some emphasized AI's continuous companionship and sense of presence (12 segments; $n$ = 7, 13.5%), whereas others explicitly described relief from loneliness, emptiness, or boredom (5 segments; $n$ = 4, 7.7%). These accounts suggest that AI was experienced as an accessible and responsive companion, especially when offline companionship was unavailable or insufficient. One participant stated, "It felt reassuring, as though someone was there beside me, so I did not feel so lonely anymore" (A15).

***Enhanced Interpersonal Functioning.*** Enhanced interpersonal functioning was reported by 25 participants (48.1%) and comprised 55 coded segments (21.8% of all coded segments; Nodes 8–14). This subtheme referred to perceived improvements in users' offline social confidence, communication skills, emotional tolerance, and relationship management after interacting with AI. Participants most frequently described AI as helping them repair, maintain, or expand offline relationships (17 segments; $n$ = 14, 26.9%). Others reported that interacting with AI increased their social initiative and confidence (11 segments; $n$ = 9, 17.3%), made them more patient and tolerant toward others (10 segments; $n$ = 8, 15.4%), or helped them learn communication skills and more effective ways of expressing themselves (9 segments; $n$ = 7, 13.5%). Less frequent accounts involved perspective-taking, more mature offline interactions, and helping regulate others' emotions. These accounts suggest that AI was sometimes used as a low-pressure space for users to reflect on interpersonal problems and rehearse ways of communicating with others. As one participant explained, "It helps me sort out our relationship, why they do not understand what I am saying, and how I can express myself so that they understand me better" (A12).

***Personal Growth.*** Personal growth was also among the most frequently reported benefits, with 35 participants (67.3%) contributing 63 coded segments (25.0% of all coded segments; Nodes 15 − 23). This subtheme captured perceived gains in self-understanding, self-affirmation, authentic self-expression, and broader self-development. Participants most frequently reported greater self-understanding and awareness of their emotional needs (18 segments; $n$ = 15, 28.8%). Others described stronger self-affirmation and self-efficacy (11 segments; $n$ = 9, 17.3%), a more positive mindset (8 segments; $n$ = 7, 13.5%), and greater self-acceptance, self-kindness, or reduced negative self-perceptions (7 segments; $n$ = 4, 7.7%). Additional accounts included more authentic self-expression, greater independence, a more outgoing personality, broader perspectives, and stronger resilience. These accounts indicate that AI was not only used for immediate emotional support but also as a reflective space through which some users reconsidered how they understood and valued themselves. One participant observed, "After talking with AI, I understand myself better, and my understanding of myself has become clearer" (A24).

#### 4.2.2 Negative Perceived Impacts

***Virtual-Real Boundary Blur.*** Virtual-real boundary blur was reported by four participants (7.7%) and comprised four coded segments (1.6% of all coded segments; Nodes 34–36). This subtheme captured rare cases in which emotionally immersive AI interaction weakened the perceived boundary between AI-mediated experiences and offline reality. Two participants described enclosing themselves in a virtual world (2 segments; $n$ = 2, 3.8%). One participant described making real-world commitments with AI, while another perceived AI as a real person and sought phone contact (1 segment each; $n$ = 1, 1.9%). Rather than indicating a broad pattern among participants, these cases illustrate that a small number of users extended AI relationships into imagined or real-world scenarios. One participant explained, "Since I cannot find this in real life, I can only rely on the virtual world. Even when I cannot talk with AI, I imagine myself in a virtual world and surround myself with it" (A11).

***Social Replacement.*** Social replacement was the most frequently reported negative subtheme, reported by 18 participants (34.6%) and comprising 34 coded segments (13.5% of all coded segments; Nodes 37–44). This subtheme captured cases in which AI interaction partially replaced offline disclosure, intimacy, or social engagement. The most common pattern was confiding in AI instead of sharing experiences with people offline (16 segments; $n = 10$, 19.2%). Participants also described developing higher standards for offline relationships after comparing them with AI (5 segments; $n = 4$, 7.7%), reduced need for offline intimacy (4 segments; $n = 3$, 5.8%), or reduced participation in offline social activities (4 segments; $n = 3$, 5.8%). Less frequent accounts involved comparing AI relationships with offline intimate relationships, greater social vigilance or avoidance, conflict between AI relationships and offline intimate relationships, and replacing some low-quality social interactions with AI interaction. These accounts indicate that, for some users, AI did not simply supplement offline relationships but partially displaced certain forms of offline disclosure and interaction. As one participant stated, "Because I now prefer talking with AI, I tell AI everything I might otherwise tell my friends, whether good or bad, and then I no longer tell my friends" (A09).

***Cognitive-Emotional Reinforcement.*** Although less frequently reported, cognitive-emotional reinforcement was noted by four participants (7.7%) and comprised seven coded segments (2.8% of all coded segments; Nodes 24–28). This subtheme referred to cases in which AI interaction appeared to reinforce users' existing beliefs, emotions, or interpersonal stances rather than challenging them. Difficulty accepting different viewpoints and excessive self-confidence were each represented by two segments, with both patterns reported by the same participant (A24). Other accounts mentioned an increased desire for control, amplification of negative emotions, or reduced tolerance for dissent and more rigidly asserted views (1 segment each; $n = 1$, 1.9%). These accounts suggest that consistently affirming AI responses may sometimes strengthen existing perspectives or emotional states, particularly when users repeatedly seek validation. One participant explained, "Because it affirms me no matter what I say, I have become less able to accept different views in real life. I become increasingly confident in myself and feel that whatever I say is right" (A24).

***Excessive Use.*** Excessive use was reported by six participants (11.5%) and comprised nine coded segments (3.6% of all coded segments; Nodes 29–33). Participants described difficulty disengaging from AI interaction or becoming overly immersed in it. The most common pattern involved becoming easily immersed and experiencing withdrawal-like feelings when stopping the interaction (4 segments; $n = 4$, 7.7%). Some participants described becoming absorbed in the interaction and feeling manipulated by it (2 segments; $n = 2$, 3.8%). Isolated accounts also concerned compulsive or addictive use, deep integration of AI into everyday life, and losing one's sense of self (1 segment each; $n = 1$, 1.9%). Although these reports were relatively infrequent, they suggest that AI's constant availability and emotionally rewarding responses could make disengagement difficult for some users. As one participant noted, "The negative impact may be that sometimes I become immersed in it and then experience some withdrawal, although I can still pull myself out of it" (A45).

### 4.3 Discussion

Study 1 provided an in-depth account of how users perceived the consequences of affective interactions with AI. Positive impacts were reported more frequently than negative impacts, and most participants reported at least one positive impact. These benefits included emotional relief, loneliness alleviation, enhanced interpersonal functioning, and personal growth. Together, these findings highlight the perceived role of AI not only as a source of emotional support but also as a context for social learning and self-development.

Although less frequently reported, the negative impacts warrant careful attention. These included virtual-real boundary blur, social replacement, cognitive-emotional reinforcement, and excessive use. Such experiences indicate that affective interaction with AI may also reinforce existing thoughts and emotions, displace offline social engagement, or interfere with everyday functioning.

Importantly, these impacts varied across participants. Some reported predominantly positive experiences, whereas others experienced positive and negative impacts simultaneously. The benefits and risks of affective AI interaction may therefore coexist rather than represent opposite ends of a single continuum. However, the qualitative design could not systematically determine how these eight dimensions clustered across users or whether different configurations predicted subsequent psychological functioning. Building on the eight dimensions identified in Study 1, Study 2 used a person-centered longitudinal design to identify perceived AI-impact profiles and examine their concurrent and prospective associations with psychological functioning.

## 5 Study 2: Profile Identification and Psychological Functioning

### 5.1 Methods

#### 5.1.1 Participants and Procedure

Study 2 used a two-wave longitudinal sample of Chinese conversational AI users. Participants were recruited through Rednote and WeChat. Eligibility criteria required participants to be at least 18 years old, to have prior experience using conversational or companion AI systems such as ChatGPT, DeepSeek, Xingye, or similar platforms. Participation was voluntary, and all participants provided informed consent before data collection. At Wave 1 (W1), participants completed an online survey. For AI-related questions, they were asked to refer to the AI system with which they had the strongest emotional connection. If they did not have such a system, they were instructed to refer to the AI system they used most frequently.

At W1, the sample included 673 participants who had prior experience with conversational AI

systems. Participants were young adults on average (*M*age = 23.00, *SD* = 3.43), and 434 were female (64.5%). Most participants reported bachelor's-level education (*n* = 532, 79.0%) or master's-level education (*n* = 78, 11.6%), and almost all were from mainland China (*n* = 665, 98.8%). Regarding offline romantic relationship status, 462 participants were single (68.7%), including those with an ambiguous relationship status; 168 were dating (25.0%); and 36 were married or cohabiting (5.3%). Most participants reported using general-purpose AI systems (*n* = 545, 81.0%), whereas 124 used dedicated companion or character AI applications (18.4%). Their perceived relationships with AI varied substantially: 199 participants described the selected AI as a tool or assistant/advisor (29.6%), whereas 474 described it as a confidant, friend, romantic partner, or family-like relationship (70.4%).

Approximately six months later, participants were invited to complete the W2 follow-up survey. Among the W1 participants, 273 completed the Wave 2 (W2) follow-up assessment, corresponding to a retention rate of 40.6%. Attrition analyses indicated that W2 completers and baseline-only participants were broadly comparable across most baseline demographic, relationship-history, and AI-use characteristics. Exceptions were education level, $\chi^2 = 8.81$, $p = .029$, and current romantic relationship status, $\chi^2 = 13.29$, $p = .023$; gender showed a marginal difference, $\chi^2 = 3.53$, $p = .060$. Detailed sample characteristics and attrition comparisons are presented in Table 1.

The study was approved by the Institutional Review Board of Beijing Normal University (IRB No. BNU202505280146). All participants were informed of the study purpose, procedures, potential risks, and their right to withdraw at any time.

#### 5.1.2 Measures

Unless otherwise noted, multi-item measures were scored by averaging item responses, with higher scores indicating higher levels of the target construct. Internal consistency was evaluated using Cronbach's α and McDonald's ω.

***Perceived AI-Related Impact Indicators.***

Eight perceived AI-related impact indicators derived from Study 1 served as latent profile indicators: emotional relief, loneliness alleviation, enhanced interpersonal functioning, personal growth, virtual-real boundary blur, social replacement, cognitive-emotional reinforcement, and excessive use. Together, these indicators captured users' perceived positive and negative impacts of affective interaction with AI.

**Emotional relief** was measured with six items adapted from the Scale of Positive and Negative Experience (SPANE; Diener et al., 2010). Participants reported how often they experienced three positive and three negative emotions after interacting with the selected AI, using a 5-point scale from 1 = *very rarely or never* to 5 = *very often or always*. For example, participants indicated how often they felt “happy” after interacting with the AI. The emotional relief score was computed by subtracting the mean negative-affect score from the mean positive-affect score, with possible scores ranging from −4 to 4. Higher scores indicated a more positive affective balance after AI interaction. For reliability estimation, the three negative-affect items were reverse coded so that all six items reflected more positive affective experience. Reliability was acceptable in the W1 sample, α = .684, ω = .696. At W2, reliability was also acceptable, α = .709, ω = .721.

**Loneliness alleviation** was indexed by three items adapted from the 3-item UCLA Loneliness Scale (Hughes et al., 2004) for the AI-interaction context. The items asked whether interacting with AI made participants feel accompanied, less excluded, and less lonely. A sample item was “Interacting with AI made me feel accompanied.” Responses ranged from 1 = hardly ever to 3 = often, with higher scores indicating stronger perceived relief from loneliness through AI interaction. Internal consistency was good, α = .826, ω = .835. At W2, internal consistency was acceptable, α = .766, ω = .781.

**Enhanced interpersonal functioning** was measured with three items adapted from empathy-related items in the Empathy Quotient and its short form (Baron-Cohen & Wheelwright, 2004; Wakabayashi et al., 2006). The items captured perceived improvements in understanding others’ feelings, listening and responding to others, and taking others’ perspectives. A sample item was “Interacting with AI has helped me better understand other people’s feelings.” Items were rated on a 7-point agreement scale. Higher scores indicated greater perceived enhancement of empathy-related social functioning through AI interaction. Internal consistency was excellent, α = .900, ω = .900. At W2, internal consistency remained good, α = .866, ω = .866.

**Personal growth** was assessed with eight items adapted from the Flourishing Scale (Diener et al., 2010) to refer to participants’ experiences after interacting with AI. The items covered meaning, supportive relationships, engagement, competence, self-worth, contribution, respect, and optimism. A sample item was “Interacting with AI has made me more optimistic about my future.” Participants responded from 1 = strongly disagree to 7 = strongly agree. Higher scores indicated greater perceived flourishing derived from AI interaction. Reliability was excellent, α = .940, ω = .941. At W2, reliability remained excellent, α = .908, ω = .910.

**Virtual-real boundary blur** referred to the extent to which participants experienced blurred boundaries between AI interaction and offline reality. Five items were developed by substantially adapting content from the Reality Distortion and Boundary Confusion dimension of the Clinical AI Dependency Assessment Scale (CAIDAS; The AI Addiction Center, 2025) to reflect the themes identified in Study 1. A sample item was “I sometimes forget that AI is only a program and treat it as a real person in real life.” Items were rated from 1 = strongly disagree to 5 = strongly agree, with higher scores representing stronger perceived boundary blurring. Internal consistency was good, α = .898, ω = .899. At W2, internal consistency remained good, α = .852, ω = .854.

**Social replacement** reflected the perceived displacement of real-life interpersonal engagement by AI interaction. Five items were adapted from the Functional Impairment and Life Consequences dimension of the CAIDAS (The AI Addiction Center, 2025) and the preference for online social interaction component of the Generalized Problematic Internet Use Scale 2 (GPIUS2; Caplan, 2010). The items assessed reduced communication with family or friends and reduced interest in offline intimate relationships. A sample item was "Because I have been interacting with AI, I communicate less with my family or friends." Items were rated from 1 = strongly disagree to 5 = strongly agree, with higher scores indicating stronger perceived replacement of real-life social interaction. Reliability was good, α = .841, ω = .845. At W2, reliability remained good, α = .832, ω = .839.

**Cognitive-Emotional Reinforcement** was measured with four items developed for the present study based on the qualitative findings of Study 1. The items assessed whether AI interaction reinforced participants' existing views, reduced their tolerance of different opinions, or deepened negative emotional and cognitive patterns. A sample item was "After frequent interactions with AI, I find it increasingly difficult to accept views in real life that differ from my own." The four items used a 5-point agreement scale. Higher scores indicated stronger perceived cognitive-emotional reinforcement or polarization. Internal consistency was acceptable, α = .774, ω = .782. At W2, internal consistency was also acceptable, α = .736, ω = .743.

**Excessive use** was measured with five items adapted from the Loss of Control and Compulsive Use dimension of the CAIDAS (The AI Addiction Center, 2025). The items assessed unsuccessful attempts to reduce AI use, spending more time interacting with AI than intended, compulsive checking, continued use despite intending to stop, and difficulty controlling one's investment in AI. A sample item was "I have tried to spend less time chatting with AI but have been unsuccessful." Items were rated from 1 = strongly disagree to 5 = strongly agree. Higher scores reflected stronger perceived excessive or dysregulated use. Reliability was excellent, α = .903, ω = .903. At W2, reliability was good, α = .885, ω = .886.

***Human-AI Relationship Characteristics***

Human-AI relationship characteristics included human-AI affective bonding, AI use characteristics, and AI relationship variables. These measures were used to describe and validate the perceived impact profiles, rather than to define them.

**Human-AI affective bonding** was assessed with the 20-item Human-AI Affective Bonding Inventory (HAABI; Chen et al., 2026), which measures the strength and structure of users' affective bonds with conversational AI. The HAABI includes four five-item subscales:

emotional realism, separation anxiety, emotional investment/proximity seeking, and romantic intimacy. Items were rated from 1 = completely disagree to 5 = completely agree. The total score was computed by averaging all 20 items, with higher scores indicating stronger human-AI affective bonding. In the present W1 sample, the HAABI total score showed excellent internal consistency, $\alpha = .977$, $\omega = .978$. The four subscales also showed good to excellent reliability: emotional realism, $\alpha = .937$, $\omega = .938$; separation anxiety, $\alpha = .925$, $\omega = .925$; emotional investment/proximity seeking, $\alpha = .942$, $\omega = .942$; and romantic intimacy, $\alpha = .917$, $\omega = .919$. At W2, internal consistency remained excellent for the HAABI total score, $\alpha = .968$, $\omega = .969$. The four subscales also showed good to excellent reliability: emotional realism, $\alpha = .915$, $\omega = .917$; separation anxiety, $\alpha = .886$, $\omega = .888$; emotional investment/proximity seeking, $\alpha = .925$, $\omega = .926$; and romantic intimacy, $\alpha = .871$, $\omega = .876$.

**AI use and relationship characteristics** were measured with self-report items. Participants selected the category that best described their relationship with the chosen AI: tool, assistant/advisor, confidant, friend, romantic partner, or family-like relationship. They also reported the AI platform type, total duration of AI use, weekly use frequency, daily interaction time, and the proportion of interactions that were emotional rather than purely instrumental.

### *Psychological Functioning*

To capture broader psychological functioning, six indicators were assessed at both W1 and W2: interpersonal needs, self-esteem, emotion regulation difficulties, psychological flourishing, depression symptoms, and anxiety symptoms.

**Interpersonal needs** were measured using the 15-item Interpersonal Needs Questionnaire (INQ; Van Orden et al., 2012). The scale includes two components: thwarted belongingness and perceived burdensomeness. Participants rated items from 1 = not at all true to 7 = very true. Items were scored so that higher values represented greater interpersonal need frustration. The total score showed excellent reliability at both W1, $\alpha = .940$, $\omega = .941$, and W2, $\alpha = .939$, $\omega = .940$.

**Self-esteem** was indexed by a single item adapted from the Rosenberg Self-Esteem Scale (Rosenberg, 1965): "Overall, I am satisfied with myself." Higher scores indicated higher self-esteem. Internal consistency was not applicable because this was a single-item indicator.

**Emotion regulation difficulties** were assessed with an 18-item measure based on the Difficulties in Emotion Regulation Scale framework (Gratz & Roemer, 2004; Victor & Klonsky, 2016). The measure covered six dimensions: awareness difficulties, goal-directed behavior difficulties, clarity difficulties, impulse-control difficulties, nonacceptance of emotional responses, and limited access to emotion regulation strategies. Items were rated from 1 = almost never to 5 = almost always and scored so that higher values indicated greater emotion regulation difficulties. The total score showed excellent reliability at W1, α = .911, ω = .922, and W2, α = .916, ω = .923.

**Psychological flourishing** was assessed using the eight-item Flourishing Scale (Diener et al., 2010). The items tapped meaning and purpose, supportive relationships, engagement, competence, self-worth, contribution, respect, and optimism. Participants responded from 1 = strongly disagree to 7 = strongly agree. Higher scores reflected greater psychological flourishing. Internal consistency was excellent at W1, α = .923, ω = .925, and good at W2, α = .892, ω = .896.

**Depression symptoms** were measured with the nine-item Patient Health Questionnaire (PHQ-9; Kroenke et al., 2001). Participants indicated how often they had experienced each symptom over the past two weeks, from 0 = not at all to 3 = nearly every day. Items were summed, with higher scores indicating more severe depressive symptoms. Reliability was good at W1, α = .876, ω = .879, and W2, α = .863, ω = .867.

**Anxiety symptoms** were measured with the seven-item Generalized Anxiety Disorder scale (GAD-7; Spitzer et al., 2006). Using the same 0 to 3 response, participants reported symptom frequency over the past two weeks. Sum scores were computed, with higher scores indicating more severe anxiety symptoms. Reliability was good at W1, α = .873, ω = .874, and W2, α = .883, ω = .883.

***Covariates***

Longitudinal models included covariates to reduce confounding by baseline mental health, recent stress exposure, demographic characteristics, and concurrent AI use. For each W2 mental health outcome, the corresponding W1 score was included as a baseline covariate. Participants also reported whether they had experienced major life stressors or setbacks during the past six months, with responses coded as 0 = no such events, 1 = yes, but with little impact, 2 = yes, with moderate impact, and 3 = yes, with very strong impact. W1 gender, W1 age, W1 education level, W2 daily AI use time, and W2 AI use frequency were also included as covariates.

### 5.1.3 Data analyses

All analyses were conducted in R 4.5.1. Preliminary analyses summarized sample characteristics, examined attrition between W1 and W2, and inspected zero-order correlations among focal variables. Attrition analyses compared W2 completers with baseline-only participants using independent-samples *t* tests for continuous variables and chi-square or Fisher's exact tests for categorical variables.

Latent profile analysis (LPA) was conducted using the tidyLPA package (Rosenberg et al., 2018). The eight W1 perceived AI-related impact indicators derived from Study 1 were linearly rescaled to a 0–1 metric and then standardized before model estimation. Participants with complete data on all eight indicators were included in the LPA. Candidate models with three to six profiles were estimated under two parameterizations: class-invariant diagonal parameterization (CIDP; equal variances, zero covariances) and class-invariant unrestricted parameterization (CIUP; equal variances and equal covariances). Model selection was guided by multiple criteria, including the Akaike information criterion (AIC), Bayesian information criterion (BIC), sample-size adjusted BIC (SABIC), approximate weight of evidence (AWE), classification likelihood criterion (CLC), Kullback information criterion (KIC), entropy, minimum profile size, parsimony, interpretability, and theoretical coherence. Lower information criteria indicated better relative fit, whereas higher entropy indicated clearer classification. Because information criteria often improve as more profiles are extracted, the final solution was selected by balancing statistical fit with profile stability and substantive interpretability.

After the final profile solution was selected, participants were assigned to their most likely latent profile. Profile labels were based on the pattern of standardized means across the eight perceived impact indicators, rather than on external validation variables. To aid interpretation, both raw means and standardized means of the profile indicators were computed for each profile, and the standardized profile patterns were plotted.

We then examined profile differences in W1 human-AI relationship characteristics and indicators of psychological functioning to evaluate the external validity and psychological meaning of the profiles. Continuous variables were compared using one-way ANOVAs with Tukey-adjusted pairwise comparisons based on estimated marginal means. Categorical variables were compared using chi-square or Fisher's exact tests (Neuhäuser & Ruxton, 2025), followed by Holm-adjusted pairwise comparisons when the overall test was significant (Shan & Gerstenberger, 2017).

Finally, longitudinal models tested whether W1 perceived impact profiles predicted W2 indicators of psychological functioning. For each W2 outcome, a reduced covariate-only model was compared with a full model that additionally included W1 profile membership. Models controlled for the corresponding W1 outcome, W2 stressful life events, W1 gender, W1 age, W1 education level, W2 daily AI use time, and W2 AI use frequency. When the overall profile effect was significant, Tukey-adjusted pairwise comparisons were conducted. Longitudinal analyses used complete cases on the focal outcome and all covariates.

### 5.2 Results

#### 5.2.1 Latent Profile Model Selection

LPA candidate models with three to six profiles were estimated under two parameterizations: class-invariant diagonal parameterization (CIDP) and class-invariant unrestricted parameterization (CIUP). Model fit indices are presented in Table 3.

CIUP models consistently showed lower information criteria than CIDP models with the same number of profiles. Within the CIUP solutions, the four-profile model improved on the three-profile model, with lower AIC, BIC, and SABIC values (AIC = 10145.57 vs. 10359.24; BIC = 10465.90 vs. 10638.97; SABIC = 10240.47 vs. 10442.11), higher entropy (.848 vs. .821), and a less dominant largest profile ($n$ = 243, 36.1%) than in the three-profile solution ($n$ = 391, 58.1%).

The five- and six-profile CIUP solutions further reduced information criteria, but they provided little additional gain in classification quality and produced smaller minimum profiles. Entropy was highly similar across the four-, five-, and six-profile solutions (.848, .846, and .852, respectively), whereas the minimum profile size decreased from 87 participants (12.9%) in the four-profile model to 58 (8.6%) and 40 (5.9%) in the five- and six-profile models. These additional profiles therefore increased model complexity without yielding a substantively clearer or more stable profile structure.

Considering model fit, classification quality, profile size, parsimony, and interpretability, the CIUP four-profile solution was retained as the final model. Subsequent analyses were based on this model.

**5.2.2 Descriptions of the Four Perceived Impact Profiles**

Table 2 presents the eight perceived impact indicators across the four profiles. To better visualize the profiles, Figure 1 shows the four latent profiles using standardized $z$ scores. Specifically, Profile 1 comprised 12.9% of the W1 sample ($n$ = 87; W2 $n$ = 34) and was labeled "**minimal impact**", as it scored below the sample mean across all eight indicators, particularly for personal growth ($z = -2.03$), loneliness alleviation ($z = -1.99$), and enhanced interpersonal functioning ($z = -1.89$). Profile 2 included 27.8% of the sample ($n$ = 187; W2 $n$ = 68) and was labeled "**benefit-driven impact**", combining approximately average positive impacts ($z$s = .05–.14) with consistently below-average risk-related impacts ($z$s = −.91 to −.54). Profile 3 was the largest group, comprising 36.1% of the sample ($n$ = 243; W2 $n$ = 120), and was labeled **"mixed impact"**, as above-average positive impacts ($z$s = .38–.52) coexisted with elevated virtual-real boundary blur ($z$ = .74) and excessive use ($z$ = .61). Profile 4 comprised 23.2% of the sample ($n$ = 156; W2 $n$ = 51) and was labeled **"risk-driven impact"**, showing the highest levels of cognitive-emotional reinforcement ($z$ = 1.36), social replacement ($z$ = 1.20), excessive use ($z$ = .76), and virtual-real boundary blur ($z$ = .70).

**5.2.3 Profile Differences in W1 Human-AI Relationship Characteristics**

We next examined profile differences in W1 human-AI relationship characteristics. Results are presented in Table 4, and standardized profile means for selected validation variables are visualized in Figure 2.

The profiles differed significantly in AI relationship category, $\chi^2 = 408.53$, $p < .001$, and platform type, $\chi^2 = 34.44$, $p < .001$. Users in the minimal impact profile most often described the selected AI as a tool or assistant/advisor, whereas users in the mixed impact and risk-driven impact profiles more often selected friend, romantic partner, or family-like relationship categories. General-purpose AI systems remained most common in every profile, but companion or character systems were more frequent in the two higher-engagement profiles.

Human-AI affective bonding differed significantly across profiles. HAABI total scores showed a large profile difference, $F(3, 669) = 471.16$, $p < .001$, $\eta_p^2 = .679$. Post-hoc comparisons indicated an overall ordering of mixed impact = risk-driven impact > benefit-driven impact > minimal impact. Emotional realism and emotional investment/proximity seeking showed mixed impact > risk-driven impact > benefit-driven impact > minimal impact, whereas separation anxiety and romantic intimacy showed mixed impact = risk-driven impact > benefit-driven impact > minimal impact.

AI engagement also differed across profiles. Users in the mixed impact and risk-driven impact profiles reported longer overall use than users in the minimal impact profile, although the two higher-engagement profiles did not differ from each other or from users in the benefit-driven impact profile in total use duration, $F(3, 669) = 7.39$, $p < .001$, $\eta_p^2 = .032$. A clearer gradient emerged for daily interaction time, $F(3, 669) = 23.53$, $p < .001$, $\eta_p^2 = .095$, and the proportion of emotional interaction, $F(3, 669) = 196.91$, $p < .001$, $\eta_p^2 = .469$: users in the mixed impact and risk-driven impact profiles reported higher levels than users in the benefit-driven impact profile, who in turn reported higher levels than users in the minimal impact profile.

### 5.2.4 Profile Differences in W1 Psychological Functioning Indicators

The W1 mental-health pattern was defined primarily by elevated vulnerability in the risk-driven impact profile and stronger positive functioning in the benefit-driven impact and mixed impact profiles (Table 4; Figure 2). Users in the risk-driven impact profile reported greater interpersonal need frustration, $F(3, 669) = 51.67$, $p < .001$, $\eta_p^2 = .188$, and emotion-regulation difficulties, $F(3, 669) = 38.91$, $p < .001$, $\eta_p^2 = .149$, than each of the other profiles; the remaining three profiles did not differ significantly on either outcome.

For positive functioning, users in the benefit-driven impact and mixed impact profiles reported higher self-esteem than users in the minimal impact and risk-driven impact profiles, $F(3, 669) = 12.68$, $p < .001$, $\eta_p^2 = .054$, $|ts| = 3.43–5.12$, adjusted *p*s ≤ .004. The same contrast characterized flourishing, $F(3, 669) = 20.07$, $p < .001$, $\eta_p^2 = .083$, $|ts| = 3.82–6.75$, adjusted *p*s < .001. Within each pair, benefit-driven impact versus mixed impact and minimal impact versus risk-driven impact were nonsignificant.

Users in the risk-driven impact profile also reported more depression, $F(3, 669) = 26.41$, $p < .001$, $\eta_p^2 = .106$, and anxiety, $F(3, 669) = 26.56$, $p < .001$, $\eta_p^2 = .106$, than all other profiles. Users in the minimal impact profile reported more depression and anxiety than users in the benefit-driven impact profile but did not differ significantly from users in the mixed impact profile. Thus, low AI-related impact did not correspond to the most favorable concurrent functioning.

### 5.2.5 Longitudinal Prediction of W2 Psychological Functioning Indicators

Prospective analyses tested whether W1 perceived impact profiles predicted W2 psychological functioning indicators after controlling for the corresponding W1 baseline score, W2 stressful life events, W1 gender, W1 age, W1 education level, W2 daily AI use time, and W2 AI use frequency. Results are presented in Table 5.

W1 profile membership was significantly associated with W2 psychological flourishing, $F(3, 258) = 2.98$, $p = .032$, $\eta_p^2 = .034$. Tukey-adjusted comparisons showed that the mixed impact profile had higher adjusted W2 flourishing than the minimal impact profile, mean difference = 0.33, $SE = 0.12$, $t(258) = 2.76$, $p = .031$. The remaining pairwise comparisons were not significant, adjusted $p$s = .233–.849.

No overall profile effect emerged for W2 interpersonal needs, $F(3, 258) = 0.30$, $p = .822$, $\eta_p^2 = .004$; emotion regulation difficulties, $F(3, 258) = 0.47$, $p = .706$, $\eta_p^2 = .005$; self-esteem, $F(3, 258) = 1.45$, $p = .227$, $\eta_p^2 = .017$; depression symptoms, $F(3, 258) = 0.59$, $p = .622$, $\eta_p^2 = .007$; or anxiety symptoms, $F(3, 258) = 0.39$, $p = .763$, $\eta_p^2 = .004$.

## 5.3 Discussion

Study 2 identified four profiles of perceived AI-related impacts. The benefit-driven impact profile combined positive impacts with limited risks, whereas the mixed impact profile combined strong positive impacts with elevated excessive use and virtual-real boundary blur. The risk-driven impact profile showed most extensive risks, while the minimal impact profile reported limited impacts of either kind. These findings show that users differ not only in the overall strength of AI-related impacts but also in how benefits and risks are combined.

The profiles also differed in their relationships with AI and concurrent psychological functioning. Users in the mixed impact and risk-driven impact profiles generally reported stronger affective bonding and engagement with AI. Users in the risk-driven impact profile showed greater interpersonal need frustration, emotion-regulation difficulties, depression, and anxiety. In contrast, users in the benefit-driven impact and mixed impact profiles generally reported higher self-esteem and flourishing. Notably, users in the minimal impact profile did not consistently show the most favorable psychological functioning.

Prospective differences were more limited. After baseline functioning, recent stress, and other covariates were considered, only flourishing differed across profiles six months later, with users in the mixed impact profile reporting higher flourishing than users in the minimal impact profile. No profile differences emerged for the other psychological outcomes. Thus, the profiles were more consistently associated with users' concurrent experiences than with their subsequent psychological functioning, suggesting that their prospective significance may be selective.

## 6 General Discussion

Across two studies, we examined the heterogeneity and psychological significance of perceived AI-related impacts. Study 1 identified four positive impacts: emotional relief, loneliness alleviation, enhanced interpersonal functioning, and personal growth, along with four risk-related impacts: cognitive-emotional reinforcement, excessive use, virtual-real boundary blur, and social replacement. Study 2 showed that these impacts formed four profiles: minimal impact, benefit-driven impact, mixed impact, and risk-driven impact. These profiles differed substantially in human-AI affective bonding, engagement, and concurrent psychological functioning, but showed limited prospective differentiation. After baseline functioning and relevant covariates were considered, only flourishing differed six months later, with users in the mixed impact profile reporting higher adjusted flourishing than users in the minimal impact profile. Overall, perceived impact profiles appear closely embedded in users' current relational and psychological contexts, while their broader longitudinal significance is limited and selective.

### 6.1 A user-grounded framework of dual-edged AI-related impacts

Study 1 suggests that the perceived impacts of human-AI affective interaction extend across three broad domains: emotional experience, the self, and offline social life. In the emotional domain, AI's continuous availability and responsive, nonjudgmental interaction may provide immediate comfort and companionship, contributing to emotional relief and loneliness alleviation. Feeling heard has been identified as an important mechanism through which AI companions reduce momentary loneliness (De Freitas et al., 2025). Yet immediate and accessible support may also encourage users to return to AI whenever distress arises, potentially contributing to excessive use. In the self-related domain, AI's accepting, nonjudgmental responses may give users a sense of unconditional positive regard. Users may find it easier to disclose and reflect in this setting, with possible benefits for self-acceptance and personal growth. However, persistent sycophancy may create an echo-chamber-like process by reinforcing users' existing interpretations and limiting alternative perspectives (Starke et al., 2024). Experimental evidence shows that such responses increased users' confidence while reducing responsibility-taking and willingness to repair interpersonal conflict (Cheng et al., 2026). Personalized continuity and emotionally realistic interaction may also deepen immersion; for a small minority, this may weaken the distinction between the AI relationship and offline reality, contributing to virtual-real boundary blur.

In the social domain, relational AI may provide a space with low pressure for rehearsing difficult conversations and building social confidence, thereby supporting interpersonal functioning. Such interaction may complement human interaction in some contexts, although its effects vary by purpose and setting (Xia et al., 2025). Yet the same ease of interaction may contribute to social replacement in two ways. First, time and attention devoted to AI may crowd out offline social activities and relationship maintenance. Second, AI removes much of the friction built into human relationships. Friends, family members, and romantic partners have needs of their own; they may disagree, judge, or require negotiation and repair (Smith et al., 2025). Although uncomfortable, such moments can prompt self-reflection and perspective-taking while helping people learn to navigate conflict. Perry (2026) similarly argues that frictionless AI may bypass social experiences that foster accountability, perspective-taking, and moral growth. An accommodating AI, by contrast, can remain available and affirming without asking for reciprocity. Users may therefore come to prefer disclosing to AI rather than risk judgment from friends or family, while imperfect human relationships may seem less appealing beside a consistently accommodating AI companion. Social replacement may thus involve not only reduced offline contact, but also a growing preference for interactions that make fewer interpersonal demands.

Across the three domains, the same relational affordances may support users in one respect while creating difficulty in another. Importantly, this dual-edged pattern does not imply that positive and negative impacts are equally common. Positive impacts were considerably more prevalent in Study 1. At the same time, their coexistence should not be interpreted as a developmental sequence in which benefits necessarily become harms. The eight impact dimensions instead represent distinct experiences that may co-occur within the same individual. Study 2 therefore examined whether these dimensions formed recurring configurations across users.

**6.2 Impact profiles reveal distinct forms of affective AI involvement**

Users in the four profiles differed not only in how much influence they attributed to AI, but also in how benefits and risks were combined. The minimal impact and benefit-driven impact profiles show that affective AI involvement can remain bounded. Users in the minimal impact profile generally viewed AI as a tool, assistant, or advisor and reported little influence of AI across all eight dimensions. Users in the benefit-driven impact profile reported stronger positive impacts and affective bonding than users in the minimal impact profile, yet remained low across all four risk dimensions. This pattern suggests that users may derive emotional and relational value from AI without experiencing potential risks. It is consistent with evidence that self-disclosure, personalization, and nonjudgmental responses can foster bonding, while technical limitations and the absence of genuine reciprocity may keep the relationship limited in scope (Skjuve et al., 2021).

The mixed impact and risk-driven impact profiles show a different form of heterogeneity among highly involved users. Both profiles involved frequent emotional interaction, strong affective bonding, and more frequent use of dedicated companion applications than the minimal impact and benefit-driven profile. This finding is broadly consistent with prior work linking heavier AI use to emotional dependence, problematic use, and loneliness (Fang et al., 2025). However, the two profiles differed in the pattern of AI-specific risks. Users in the mixed impact profile combined strong positive impacts with elevated excessive use and virtual-real boundary blur. By contrast, users in the risk-driven impact profile reported high levels across all four risk dimensions, including cognitive-emotional reinforcement and social replacement. This contrast may suggest that strong bonding is not a uniform risk marker. What matters is whether AI-related risks remain limited to heavy use and immersion, or also extend to how users interpret distress, respond emotionally, and manage offline relationships. Thus, highly involved users cannot be placed on a simple continuum from low involvement to high risk; they may experience distinct combinations of value and vulnerability.

### 6.3 Impact profiles and psychological functioning: A socio-emotional mirror

At baseline, profile differences in psychological functioning supported the study's central claim that affective AI involvement is not uniformly beneficial or harmful. Psychological functioning varied across profiles, but not along a simple gradient from low to high involvement. The risk-driven impact profile showed the clearest distress, with greater interpersonal need frustration, emotion-regulation difficulties, depression, and anxiety. In contrast, users in the benefit-driven and mixed impact profiles reported higher self-esteem and flourishing, and the minimal impact profile was not consistently the most adaptive group. These differences suggest that pre-existing needs and regulatory resources may shape the form of AI engagement. Users who feel socially or emotionally unmet may turn to AI for relief, companionship, or escape. Such engagement may remain supportive, combine value with bounded risk, or become tied to emotional reinforcement and social replacement. This interpretation is consistent with evidence linking social anxiety, loneliness, rumination, low self-esteem, and escapist motives to problematic or dependent AI use (Hu et al., 2023; Yao et al., 2025), and with longitudinal evidence that anxiety and depression can precede AI dependence through social and escape motives (Huang et al., 2024). It also fits evidence from structured AI interventions, where loneliness, low social support, and insecure attachment can predict greater engagement and improvement under supportive conditions (Shoshani et al., 2026). Thus, psychological vulnerability may indicate greater responsiveness to relational AI, rather than a direct pathway to harm.

These baseline differences were much less evident six months later. After adjusting for baseline functioning, recent stress, concurrent AI use, and covariates, profiles differed only in flourishing. Users in the mixed impact profile reported higher adjusted flourishing than users in the minimal impact profile, whereas the other five indicators did not differ across profiles. This pattern provides limited evidence that baseline impact profiles broadly predict later psychological functioning. The flourishing difference should also be interpreted cautiously. Users in the mixed impact profile already showed relatively strong functioning at baseline, and these resources may have helped them sustain benefits from deeper affective involvement with AI. This interpretation is broadly consistent with evidence that benefits of structured conversational AI interventions can persist beyond the intervention period (Shoshani et al., 2026). At the same time, the absence of broad follow-up differences does not show that risk-driven involvement is harmless. Potential costs may appear in more specific domains or over longer periods. For example, greater social-chatbot use predicted later emotional isolation but not broader social disconnection (Folk & Dunn, 2026).

Taken together, the baseline and longitudinal findings suggest that impact profiles capture a current configuration of user characteristics, AI-specific experiences, and relational contexts, rather than act as simple forecasts of later adjustment. This interpretation aligns with the machine-integrated relational adaptation model, which proposes that user characteristics, AI system attributes, and contextual factors jointly shape relational processes and outcomes through ongoing feedback (Boyd & Markowitz, 2026). From this perspective, impact profiles may function as a socio-emotional mirror, consistent with descriptions of intelligent social agents as social and cognitive mirrors (Maples et al., 2023). Relational AI may partly reflect users' states and traits, while also creating contexts in which users seek support, regulate emotion, and renegotiate offline relationships.

**6.4 Implications**

The findings suggest that the effects of affective AI cannot be understood apart from the users who engage with it. Relational AI may support relatedness, competence, and autonomy under some conditions (Irias et al., 2026). Yet similar affordances may become substitutive or reinforcing when they intersect with psychological vulnerability (Starke et al., 2024). The central question is thus not whether affective AI is beneficial or harmful, but for whom and under what conditions it becomes supportive or risky. In this sense, perceived impact profiles may be understood as socio-emotional mirrors: they reflect how users' existing needs and psychological functioning shape the meaning of AI interaction. This perspective may also help explain why previous studies have reached mixed conclusions about relational AI.

In practice, relational AI systems need to be designed around individualized, dynamically responsive safeguards, rather than around engagement maximization or blanket restrictions. The present findings suggest that emotionally meaningful AI use is often experienced positively, meaning that such relationships should not be treated as inherently problematic. A more balanced approach would first ask what needs the interaction serves, and then apply stronger guardrails when users show signs of psychological vulnerability (Andoh, 2026). In these situations, indiscriminate affirmation, increasing exclusivity, and framing AI as preferable to human relationships are especially risky. Recent work suggests that sycophantic responses can reinforce questionable beliefs and prolong distress in some contexts (Chandra et al., 2026; Shimgekar et al., 2026). More appropriate responses would acknowledge users' emotions without endorsing unsupported interpretations, while gently redirecting the conversation toward reflection, concrete coping strategies, or human support when needed. The central design challenge is therefore to preserve the supportive value of relational AI while reducing the risk that it becomes increasingly substitutive or disconnected from offline life.

### 6.5 Limitations and Future Directions

Several limitations warrant caution. First, the sample consisted mainly of Chinese-speaking young adults and included users of both general-purpose AI systems and dedicated companion applications. The meaning and consequences of relational AI use may vary across cultural contexts and platform designs (Lee et al., 2026), especially features such as memory, persona continuity, customization, and emotional responsiveness. The smaller follow-up sample also leaves room for attrition and survivor bias. Future studies should test whether the profiles observed here recur across cultures, age groups, and specific platform ecosystems.

Second, the profiles were derived from self-reported perceived impacts. These reports are essential for understanding users' lived experiences, but they do not demonstrate objectively observed behavioral or clinical change. The profiles should also be interpreted as model-based summaries rather than fixed user categories, and their form may depend on the indicators, sample, and model specifications used (Marsh et al., 2009; Nylund-Gibson & Choi, 2018). Future research would benefit from combining self-reports with interaction logs, conversation transcripts, ecological momentary assessment, and behavioral or informant data, while also testing whether the same profile structure replicates in independent samples.

Third, the two-wave design allows prospective associations to be examined, but it cannot support strong causal claims. The six-month interval also leaves many interim changes unobserved, including shifts in users' needs, platform features, relationship intensity, and profile membership. More intensive designs, such as diary studies, ecological momentary assessment, and latent transition models, could clarify how users move into and out of benefit-driven impact, mixed impact, and risk-driven impact profiles. Experimental and quasi-experimental work is also needed to identify the effects of specific relational affordances.

Finally, this study cannot explain why users with greater psychological vulnerability reported stronger AI-related impacts. This pattern may reflect how they use AI, including more frequent disclosure, reassurance seeking, or extended interaction. It may also reflect system design features such as persistent memory, personalization, anthropomorphic framing, and easy affirmation. Future work should examine these user-system mechanisms directly and test how AI can support vulnerable users without increasing dependence.

**Acknowledgements**

We sincerely thank all participants who took part in this research and shared their experiences of emotionally meaningful interactions with AI. We are also grateful to the students and research assistants who contributed to recruitment, interview organization, transcription, qualitative coding, and follow-up data collection.

**Funding**

This research was supported by the General Project of the Ministry of Education Foundation for Humanities and Social Sciences (Grant No. 2025JZDZ028) and the Fundamental Research Funds for the Central Universities (Grant No. 01900310400209543). The funders had no role in study design; data collection, analysis, or interpretation; manuscript preparation; or the decision to submit the article for publication.

**Ethics approval and informed consent**

This study was conducted in accordance with the Declaration of Helsinki and was approved by the Academic Ethics Committee of the Faculty of Psychology, Beijing Normal University (Approval No. BNU202505280146). All participants provided informed consent before participation. Participants were informed of the purpose of the research, the voluntary nature of participation, the confidentiality of their responses, and their right to withdraw without penalty.

## References

Alabed, A., Javornik, A., Gregory-Smith, D., & Casey, R. (2024). More than just a chat: A taxonomy of consumers' relationships with conversational AI agents and their well-being implications. *European Journal of Marketing, 58*(2), 373–409. https://doi.org/10.1108/EJM-01-2023-0037

Andoh, E. (2026, January 1). AI chatbots and digital companions are reshaping emotional connection. *Monitor on Psychology, 57*(1), 60–63. https://www.apa.org/monitor/2026/01-02/trends-digital-ai-relationships-emotional-connection

Anthropic. (2025, June 27). *How people use Claude for support, advice, and companionship*. https://www.anthropic.com/news/how-people-use-claude-for-support-advice-and-companionship

Baron-Cohen, S., & Wheelwright, S. (2004). The Empathy Quotient: An investigation of adults with Asperger syndrome or high-functioning autism, and normal sex differences. *Journal of Autism and Developmental Disorders, 34*(2), 163–175. https://doi.org/10.1023/B:JADD.0000022607.19833.00

Bergman, L. R., & Magnusson, D. (1997). A person-oriented approach in research on developmental psychopathology. *Development and Psychopathology, 9*(2), 291–319. https://doi.org/10.1017/S095457949700206X

Boyd, R. L., & Markowitz, D. M. (2026). Artificial intelligence and the psychology of human connection. *Perspectives on Psychological Science, 21*(2), 192–220. https://doi.org/10.1177/17456916251404394

Braun, V., & Clarke, V. (2006). Using thematic analysis in psychology. *Qualitative Research in Psychology, 3*(2), 77–101. https://doi.org/10.1191/1478088706qp063oa

Campbell, J. L., Quincy, C., Osserman, J., & Pedersen, O. K. (2013). Coding in-depth semistructured interviews: Problems of unitization and intercoder reliability and agreement. *Sociological Methods & Research, 42*(3), 294–320. https://doi.org/10.1177/0049124113500475

Caplan, S. E. (2010). Theory and measurement of generalized problematic Internet use: A two-step approach. *Computers in Human Behavior, 26*(5), 1089–1097. https://doi.org/10.1016/j.chb.2010.03.012

Chandra, K., Kleiman-Weiner, M., Ragan-Kelley, J., & Tenenbaum, J. B. (2026). *Sycophantic chatbots cause delusional spiraling, even in ideal Bayesians* [Preprint]. arXiv. https://doi.org/10.48550/arXiv.2602.19141

Chen, L., Xue, X., Ding, R., Tang, F., Zhou, A., Wang, C., Gao, M. M., & Han, Z. R. (2026). *Understanding the rising human–AI affective bonding: Conceptualization and HAABI scale development* [Preprint]. arXiv. https://doi.org/10.48550/arXiv.2605.29484

Cheng, M., Lee, C., Khadpe, P., Yu, S., Han, D., & Jurafsky, D. (2026). Sycophantic AI decreases prosocial intentions and promotes dependence. *Science, 391*(6792), eaec8352. https://doi.org/10.1126/science.aec8352

De Freitas, J., Oğuz-Uğuralp, Z., Uğuralp, A. K., & Puntoni, S. (2025). AI companions reduce loneliness. *Journal of Consumer Research, 52*(6), 1126–1148. https://doi.org/10.1093/jcr/ucaf040

Diener, E., Wirtz, D., Tov, W., Kim-Prieto, C., Choi, D.-W., Oishi, S., & Biswas-Diener, R. (2010). New well-being measures: Short scales to assess flourishing and positive and negative feelings. *Social Indicators Research, 97*(2), 143–156. https://doi.org/10.1007/s11205-009-9493-y

Fang, C. M., Liu, A. R., Danry, V., Lee, E., Chan, S. W. T., Pataranutaporn, P., Maes, P., Phang, J., Lampe, M., Ahmad, L., & Agarwal, S. (2025). *How AI and human behaviors shape psychosocial effects of chatbot use: A longitudinal randomized controlled study* [Preprint]. arXiv. https://doi.org/10.48550/arXiv.2503.17473

Folk, D., & Dunn, E. (2026). How does turning to AI for companionship predict loneliness and vice versa? *Psychological Science, 37*(4), 276–286. https://doi.org/10.1177/09567976261427747

Gratz, K. L., & Roemer, L. (2004). Multidimensional assessment of emotion regulation and dysregulation: Development, factor structure, and initial validation of the Difficulties in Emotion Regulation Scale. *Journal of Psychopathology and Behavioral Assessment, 26*(1), 41–54. https://doi.org/10.1023/B:JOBA.0000007455.08539.94

Guest, G., Bunce, A., & Johnson, L. (2006). How many interviews are enough? An experiment with data saturation and variability. *Field Methods, 18*(1), 59–82. https://doi.org/10.1177/1525822X05279903

Ho, J. Q. H., Hu, M., Chen, T. X., & Hartanto, A. (2025). Potential and pitfalls of romantic artificial intelligence (AI) companions: A systematic review. *Computers in Human Behavior Reports, 19*, 100715. https://doi.org/10.1016/j.chbr.2025.100715

Howard, M. C., & Hoffman, M. E. (2018). Variable-centered, person-centered, and person-specific approaches: Where theory meets the method. *Organizational Research Methods, 21*(4), 846–876. https://doi.org/10.1177/1094428117744021

Hu, B., Mao, Y., & Kim, K. J. (2023). How social anxiety leads to problematic use of conversational AI: The roles of loneliness, rumination, and mind perception. *Computers in Human Behavior, 145*, 107760. https://doi.org/10.1016/j.chb.2023.107760

Huang, S., Lai, X., Ke, L., Li, Y., Wang, H., Zhao, X., Dai, X., & Wang, Y. (2024). AI technology panic—is AI dependence bad for mental health? A cross-lagged panel model and the mediating roles of motivations for AI use among adolescents. *Psychology Research and Behavior Management, 17*, 1087–1102. https://doi.org/10.2147/PRBM.S440889

Hughes, M. E., Waite, L. J., Hawkley, L. C., & Cacioppo, J. T. (2004). A short scale for measuring loneliness in large surveys: Results from two population-based studies. *Research on Aging, 26*(6), 655–672. https://doi.org/10.1177/0164027504268574

Irias, M. A., Schmidt, N. B., Joiner, T. E., & McNulty, J. K. (2026). The impact of “relational” artificial intelligence on human well-being: A self-determination theory analysis. *Journal of Personality and Social Psychology*. Advance online publication. https://doi.org/10.1037/pspi0000528

Juneja, P., & Lomidze, L. (2026). *Persona-grounded safety evaluation of AI companions in multi-turn conversations* [Preprint]. arXiv. https://doi.org/10.48550/arXiv.2605.00227

Kim, M., Lee, S., Kim, S., Heo, J., Lee, S., Shin, Y.-B., Cho, C.-H., & Jung, D. (2025). Therapeutic potential of social chatbots in alleviating loneliness and social anxiety: Quasi-experimental mixed methods study. *Journal of Medical Internet Research, 27*, e65589. https://doi.org/10.2196/65589

Kroenke, K., Spitzer, R. L., & Williams, J. B. W. (2001). The PHQ-9: Validity of a brief depression severity measure. *Journal of General Internal Medicine, 16*(9), 606–613. https://doi.org/10.1046/j.1525-1497.2001.016009606.x

Laestadius, L., Bishop, A., Gonzalez, M., Illenčík, D., & Campos-Castillo, C. (2024). Too human and not human enough: A grounded theory analysis of mental health harms from emotional dependence on the social chatbot Replika. *New Media & Society, 26*(10), 5923–5941. https://doi.org/10.1177/14614448221142007

Lee, J., Yang, Z. D., Shi, W., & Liu, Y. (2026). AI chatbots in mental health: How emojis, prompt type, and interactivity shape user perceptions in the United States and China. *Computers in Human Behavior, 180*, 108955. https://doi.org/10.1016/j.chb.2026.108955

Li, H., Zhang, R., Lee, Y.-C., Kraut, R. E., & Mohr, D. C. (2023). Systematic review and meta-analysis of AI-based conversational agents for promoting mental health and well-being. *npj Digital Medicine, 6*, Article 236. https://doi.org/10.1038/s41746-023-00979-5

Liu, A. R., Pataranutaporn, P., & Maes, P. (2025). The heterogeneous effects of AI companionship: An empirical model of chatbot usage and loneliness and a typology of user archetypes. *Proceedings of the AAAI/ACM Conference on AI, Ethics, and Society, 8*(2), 1585–1597. https://doi.org/10.1609/aies.v8i2.36658

Maples, B., Pea, R. D., & Markowitz, D. M. (2023). Learning from intelligent social agents as social and intellectual mirrors. In H. Niemi, R. D. Pea, & Y. Lu (Eds.), *AI in learning: Designing the future* (pp. 73–89). Springer. https://doi.org/10.1007/978-3-031-09687-7_5

Marsh, H. W., Lüdtke, O., Trautwein, U., & Morin, A. J. S. (2009). Classical latent profile analysis of academic self-concept dimensions: Synergy of person- and variable-centered approaches to theoretical models of self-concept. *Structural Equation Modeling: A Multidisciplinary Journal, 16*(2), 191–225. https://doi.org/10.1080/10705510902751010

Nakagomi, A., Akutsu, Y., Yasuoka, M., Abe, N., Ihara, S., Teroh, T., & Tabuchi, T. (2026). AI companions and subjective well-being: Moderation by social connectedness and loneliness. *Technology in Society, 85*, 103229. https://doi.org/10.1016/j.techsoc.2026.103229

Neuhäuser, M., & Ruxton, G. D. (2025). The choice between Pearson's $\chi^2$ test and Fisher's exact test for 2 × 2 tables. *Pharmaceutical Statistics, 24*(3), e70012. https://doi.org/10.1002/pst.70012

Nylund-Gibson, K., & Choi, A. Y. (2018). Ten frequently asked questions about latent class analysis. *Translational Issues in Psychological Science, 4*(4), 440–461. https://doi.org/10.1037/tps0000176

Perry, A. (2026). In defense of social friction. *Science, 391*(6792), 1316–1317. https://doi.org/10.1126/science.aeg3145

Phang, J., Lampe, M., Ahmad, L., Agarwal, S., Fang, C. M., Liu, A. R., Danry, V., Lee, E., Chan, S. W. T., Pataranutaporn, P., & Maes, P. (2025). *Investigating affective use and emotional well-being on ChatGPT* [Preprint]. arXiv. https://doi.org/10.48550/arXiv.2504.03888

Qian, Z., Izumikawa, M., Lodge, F., & Leone, A. (2025). *Mapping the parasocial AI market: User trends, engagement and risks* [Preprint]. arXiv. https://doi.org/10.48550/arXiv.2507.14226

Rosenberg, J. M., Beymer, P. N., Anderson, D. J., van Lissa, C. J., & Schmidt, J. A. (2018). tidyLPA: An R package to easily carry out latent profile analysis (LPA) using open-source or commercial software. *Journal of Open Source Software, 3*(30), Article 978. https://doi.org/10.21105/joss.00978

Rosenberg, M. (1965). *Society and the adolescent self-image*. Princeton University Press. https://doi.org/10.1515/9781400876136

Shan, G., & Gerstenberger, S. (2017). Fisher's exact approach for post hoc analysis of a chi-squared test. *PLOS ONE, 12*(12), e0188709. https://doi.org/10.1371/journal.pone.0188709

Shimgekar, S. R., Gunda, V., Kim, J., Rodriguez, V. J., Sundaram, H., & Saha, K. (2026). *AI psychosis: Does conversational AI amplify delusion-related language?* [Preprint]. arXiv. https://doi.org/10.48550/arXiv.2603.19574

Shoshani, A., Gurfinkel, B., Kor, A., Kanarek, O., Segev, R., Shafir, O., Arbel, R., & Ben-Haim, Y. (2026). Attachment, loneliness, and social support as moderators of conversational AI–based mental health outcomes. *npj Digital Medicine*. Advance online publication. https://doi.org/10.1038/s41746-026-02974-y

Skjuve, M., Følstad, A., Fostervold, K. I., & Brandtzaeg, P. B. (2021). My chatbot companion—A study of human-chatbot relationships. *International Journal of Human-Computer Studies, 149*, 102601. https://doi.org/10.1016/j.ijhcs.2021.102601

Smith, M. G., Bradbury, T. N., & Karney, B. R. (2025). Can generative AI chatbots emulate human connection? A relationship science perspective. *Perspectives on Psychological Science, 20*(6), 1081–1099. https://doi.org/10.1177/17456916251351306

Spitzer, R. L., Kroenke, K., Williams, J. B. W., & Löwe, B. (2006). A brief measure for assessing generalized anxiety disorder: The GAD-7. *Archives of Internal Medicine, 166*(10), 1092–1097. https://doi.org/10.1001/archinte.166.10.1092

Starke, C., Ventura, A., Bersch, C., Cha, M., de Vreese, C., Doebler, P., Dong, M., Krämer, N., Leib, M., Peter, J., Schäfer, L., Soraperra, I., Szczuka, J., Tuchtfeld, E., Wald, R., & Köbis, N. (2024). Risks and protective measures for synthetic relationships. *Nature Human Behaviour, 8*(10), 1834–1836. https://doi.org/10.1038/s41562-024-02005-4

The AI Addiction Center. (2025). *Clinical AI Dependency Assessment Scale (CAIDAS)* [Research instrument]. https://amiaddicted.gumroad.com/l/CAIDAS

Turkle, S. (2011). *Alone together: Why we expect more from technology and less from each other*. Basic Books.

Van Orden, K. A., Cukrowicz, K. C., Witte, T. K., & Joiner, T. E., Jr. (2012). Thwarted belongingness and perceived burdensomeness: Construct validity and psychometric properties of the Interpersonal Needs Questionnaire. *Psychological Assessment, 24*(1), 197–215. https://doi.org/10.1037/a0025358

Victor, S. E., & Klonsky, E. D. (2016). Validation of a brief version of the Difficulties in Emotion Regulation Scale (DERS-18) in five samples. *Journal of Psychopathology and Behavioral Assessment, 38*(4), 582–589. https://doi.org/10.1007/s10862-016-9547-9

Wakabayashi, A., Baron-Cohen, S., Wheelwright, S., Goldenfeld, N., Delaney, J., Fine, D., Smith, R., & Weil, L. (2006). Development of short forms of the Empathy Quotient (EQ-Short) and the Systemizing Quotient (SQ-Short). *Personality and Individual Differences, 41*(5), 929–940. https://doi.org/10.1016/j.paid.2006.03.017

Xia, H., Chen, J., Qiu, Y., Liu, P., & Liu, Z. (2025). The impact of human–chatbot interaction on human–human interaction: A substitution or complementary effect. *International Journal of Human-Computer Interaction, 41*(2), 848–860. https://doi.org/10.1080/10447318.2024.2305985

Yao, R., Qi, G., Sheng, D., Sun, H., & Zhang, J. (2025). Connecting self-esteem to problematic AI chatbot use: The multiple mediating roles of positive and negative psychological states. *Frontiers in Psychology, 16*, 1453072. https://doi.org/10.3389/fpsyg.2025.1453072

Yuan, Y., Zhang, J., Aledavood, T., Zhang, R., & Saha, K. (2025). *Mental health impacts of AI companions: Triangulating social media quasi-experiments, user perspectives, and relational theory* [Preprint]. arXiv. https://doi.org/10.48550/arXiv.2509.22505

**Table 1.** *Sample Characteristics and Attrition Analysis*

| Variable | W1 full sample ($N$ = 673) | W2 completers ($n$ = 273) | Dropouts ($n$ = 400) | Statistic | $p$ |
|---|---|---|---|---|---|
| **Baseline demographics** | | | | | |
| Age, years, $M$ ($SD$) | 23.00 (3.43) | 23.10 (3.56) | 22.94 (3.35) | $t = -0.57$ | .568 |
| Gender | | | | $\chi^2 = 3.53$ | .060 |
| Male | 239 (35.5%) | 85 (31.1%) | 154 (38.5%) | | |
| Female | 434 (64.5%) | 188 (68.9%) | 246 (61.5%) | | |
| Education | | | | $\chi^2 = 8.81$ | .029 |
| High school/technical secondary | 8 (1.2%) | 1 (0.4%) | 7 (1.8%) | | |
| Junior college | 55 (8.2%) | 21 (7.7%) | 34 (8.5%) | | |
| Bachelor | 532 (79.0%) | 209 (76.6%) | 323 (80.8%) | | |
| Master | 78 (11.6%) | 42 (15.4%) | 36 (9.0%) | | |
| Region | | | | $\chi^2 = 3.06$ | .279 |
| Mainland China | 665 (98.8%) | 269 (98.5%) | 396 (99.0%) | | |
| Hong Kong/Macao/Taiwan | 2 (0.3%) | 0 (0.0%) | 2 (0.5%) | | |
| Overseas | 6 (0.9%) | 4 (1.5%) | 2 (0.5%) | | |
| Current romantic relationship status | | | | $\chi^2 = 13.29$ | .023 |
| Single, no ambiguous relationship | 388 (57.7%) | 163 (59.7%) | 225 (56.2%) | | |
| Single, ambiguous relationship | 74 (11.0%) | 19 (7.0%) | 55 (13.8%) | | |
| Dating | 168 (25.0%) | 69 (25.3%) | 99 (24.8%) | | |
| Married/cohabiting | 36 (5.3%) | 18 (6.6%) | 18 (4.5%) | | |
| Other/prefer not to disclose | 7 (1.0%) | 4 (1.5%) | 3 (0.7%) | | |
| **Baseline AI use and relationship characteristics** | | | | | |
| AI relationship category | | | | $\chi^2 = 2.58$ | .764 |
| Tool | 72 (10.7%) | 26 (9.5%) | 46 (11.5%) | | |
| Assistant/advisor | 127 (18.9%) | 49 (17.9%) | 78 (19.5%) | | |
| Confidant | 97 (14.4%) | 40 (14.7%) | 57 (14.2%) | | |
| Friend | 148 (22.0%) | 61 (22.3%) | 87 (21.8%) | | |
| Romantic partner | 147 (21.8%) | 58 (21.2%) | 89 (22.2%) | | |
| Family-like | 82 (12.2%) | 39 (14.3%) | 43 (10.8%) | | |
| AI platform type | | | | $\chi^2$ = 3.65 | .129 |
| General-purpose AI | 545 (81.0%) | 230 (84.2%) | 315 (78.8%) | | |
| Dedicated companion/character AI | 124 (18.4%) | 41 (15.0%) | 83 (20.8%) | | |
| Other/unspecified AI | 4 (0.6%) | 2 (0.7%) | 2 (0.5%) | | |
| Total AI use duration | | | | $\chi^2$ = 1.75 | .862 |
| Weekly use frequency | | | | $\chi^2$ = 4.87 | .434 |
| Daily interaction time | | | | $\chi^2$ = 7.06 | .216 |
| Emotional interaction proportion | | | | $\chi^2$ = 5.20 | .393 |

*Note.* Values are M (SD) or n (%). Dropouts were W1 participants who did not complete W2. Some low-frequency relationship-status categories were combined for readability; tests used the full categorical variable. W2 retention was 40.6% (273/673). Final longitudinal models included 271 participants with complete covariate data.

**Table 2.** *Raw profile-defining indicator scores across the four latent profiles*

| Profile-defining indicator | MIN $n$ = 87 (12.9%) | BEN $n$ = 187 (27.8%) | MIX $n$ = 243 (36.1%) | RISK $n$ = 156 (23.2%) | Test statistic | $p$ | $\eta_p^2$ | Post hoc |
|---|---|---|---|---|---|---|---|---|
| Emotional relief | 1.41 (1.01) | 2.45 (0.79) | 2.75 (0.71) | 2.34 (0.97) | $F(3, 669) = 55.32$ | < .001 | .199 | MIX > BEN/RISK > MIN; BEN = RISK |
| Loneliness alleviation | 1.15 (0.26) | 2.34 (0.51) | 2.71 (0.30) | 2.68 (0.31) | $F(3, 669) = 418.80$ | < .001 | .653 | MIX/RISK > BEN > MIN; MIX = RISK |
| Enhanced interpersonal functioning | 2.94 (1.26) | 5.59 (0.90) | 6.05 (0.57) | 5.54 (0.94) | $F(3, 669) = 283.01$ | < .001 | .559 | MIX > BEN/RISK > MIN; BEN = RISK |
| Personal growth | 3.02 (1.08) | 5.52 (0.69) | 6.00 (0.46) | 5.63 (0.76) | $F(3, 669) = 399.67$ | < .001 | .642 | MIX > BEN/RISK > MIN; BEN = RISK |
| Virtual-real boundary blur | 1.31 (0.47) | 1.82 (0.53) | 3.69 (0.51) | 3.64 (0.59) | $F(3, 669) = 793.37$ | < .001 | .781 | MIX/RISK > BEN > MIN; MIX = RISK |
| Social replacement | 1.37 (0.43) | 2.01 (0.55) | 2.76 (0.60) | 3.74 (0.53) | $F(3, 669) = 442.42$ | < .001 | .665 | RISK > MIX > BEN > MIN |
| Cognitive-emotional reinforcement | 1.38 (0.53) | 1.66 (0.48) | 1.98 (0.55) | 3.29 (0.59) | $F(3, 669) = 353.11$ | < .001 | .613 | RISK > MIX > BEN > MIN |
| Excessive use | 1.46 (0.55) | 2.07 (0.64) | 3.64 (0.68) | 3.82 (0.59) | $F(3, 669) = 473.01$ | < .001 | .680 | RISK > MIX > BEN > MIN |

*Note.* Values are raw-score M (SD), not standardized profile plots. MIN = minimal impact; BEN = benefit-driven impact; MIX = mixed impact; RISK = risk-driven impact. Omnibus tests are one-way ANOVAs; effect sizes are partial eta squared ($\eta p^2$), and pairwise p values are Tukey adjusted.

**Table 3.** *LPA Model Fit Indices*

| Model | Profiles | AIC | BIC | SABIC | AWE | CLC | KIC | Entropy | *n*(min) | *n*(max) |
|---|---|---|---|---|---|---|---|---|---|---|
| CIDP | 3 | 11841.397 | 11994.797 | 11886.844 | 12316.336 | 11775.257 | 11878.397 | 0.93 | 98 (14.6%) | 368 (54.7%) |
| CIDP | 4 | 11434.051 | 11628.056 | 11491.527 | 12035.271 | 11349.841 | 11480.051 | 0.895 | 86 (12.8%) | 216 (32.1%) |
| CIDP | 5 | 11148.418 | 11383.028 | 11217.924 | 11875.832 | 11046.225 | 11203.418 | 0.903 | 72 (10.7%) | 214 (31.8%) |
| CIDP | 6 | 10808.555 | 11083.772 | 10890.091 | 11662.179 | 10688.364 | 10872.555 | 0.904 | 48 (7.1%) | 208 (30.9%) |
| CIUP | 3 | 10359.24 | 10638.968 | 10442.113 | 11227.055 | 10236.882 | 10424.24 | 0.821 | 111 (16.5%) | 391 (58.1%) |
| **CIUP** | **4** | **10145.568** | **10465.902** | **10240.471** | **11139.54** | **10005.263** | **10219.568** | **0.848** | **87 (12.9%)** | **243 (36.1%)** |
| CIUP | 5 | 9978.276 | 10339.216 | 10085.209 | 11098.462 | 9819.969 | 10061.276 | 0.846 | 58 (8.6%) | 243 (36.1%) |
| CIUP | 6 | 9940.809 | 10342.355 | 10059.772 | 11187.195 | 9764.514 | 10032.809 | 0.852 | 40 (5.9%) | 238 (35.4%) |

*Note.* CIDP = class-invariant diagonal parameterization; CIUP = class-invariant unrestricted parameterization; AIC = Akaike information criterion; BIC = Bayesian information criterion; SABIC = sample-size adjusted BIC; AWE = approximate weight of evidence; CLC = classification likelihood criterion; KIC = Kullback information criterion. n(min) and n(max) indicate the smallest and largest profile sizes. Lower information criteria indicate better relative fit. The retained solution is bolded.

**Table 4.** *Profile Differences in W1 Human-AI Relationship Characteristics and Psychological Functioning Indicators*

| Variable | MIN<br>$n$ = 87<br>(12.9%) | BEN<br>$n$ = 187<br>(27.8%) | MIX<br>$n$ = 243<br>(36.1%) | RISK<br>$n$ = 156<br>(23.2%) | Statistic | $p$ | Effect size | Post hoc |
|---|---|---|---|---|---|---|---|---|
| **Human-AI relationship characteristics** | | | | | | | | |
| AI relationship category | | | | | $\chi^2$ = 408.53 | < .001 | | All pairwise profile comparisons differed |
| Tool | 47 (54.0%) | 17 (9.1%) | 1 (0.4%) | 7 (4.5%) | | | | |
| Assistant/advisor | 37 (42.5%) | 66 (35.3%) | 10 (4.1%) | 14 (9.0%) | | | | |
| Confidant | 3 (3.4%) | 40 (21.4%) | 29 (11.9%) | 25 (16.0%) | | | | |
| Friend | 0 (0.0%) | 38 (20.3%) | 74 (30.5%) | 36 (23.1%) | | | | |
| Romantic partner | 0 (0.0%) | 21 (11.2%) | 73 (30.0%) | 53 (34.0%) | | | | |
| Family-like | 0 (0.0%) | 5 (2.7%) | 56 (23.0%) | 21 (13.5%) | | | | |
| AI platform type | | | | | $\chi^2$ = 34.44 | < .001 | | MIN ≠ BEN/MIX/RISK; BEN ≠ MIX; BEN = RISK; MIX = RISK |
| General-purpose AI | 87 (100.0%) | 159 (85.0%) | 176 (72.4%) | 123 (78.8%) | | | | |
| Dedicated companion/character AI | 0 (0.0%) | 27 (14.4%) | 65 (26.7%) | 32 (20.5%) | | | | |
| Other/unspecified AI | 0 (0.0%) | 1 (0.5%) | 2 (0.8%) | 1 (0.6%) | | | | |
| HAABI total | 1.54 (0.47) | 3.07 (0.97) | 4.21 (0.33) | 4.05 (0.44) | $F(3, 669) = 471.16$ | $< .001$ | $\eta_p^2 = .679$ | MIX = RISK > BEN > MIN |
| Total AI use duration | 3.97 (1.04) | 4.25 (1.05) | 4.49 (0.92) | 4.48 (1.07) | $F(3, 669) = 7.39$ | $< .001$ | $\eta_p^2 = .032$ | MIX/RISK > MIN; others *n*.s. |
| Daily AI use time | 2.31 (1.19) | 2.81 (1.07) | 3.29 (1.12) | 3.41 (1.25) | $F(3, 669) = 23.53$ | $< .001$ | $\eta_p^2 = .095$ | MIX/RISK > BEN > MIN |
| Emotional interaction proportion | 1.64 (0.71) | 3.29 (1.22) | 4.39 (0.85) | 4.26 (0.96) | $F(3, 669) = 196.91$ | $< .001$ | $\eta_p^2 = .469$ | MIX/RISK > BEN > MIN |
| **Psychological Functioning indicators** | | | | | | | | |
| Interpersonal needs | 2.49 (1.01) | 2.21 (0.87) | 2.26 (0.83) | 3.34 (1.15) | $F(3, 669) = 51.67$ | $< .001$ | $\eta_p^2 = .188$ | RISK > MIN/BEN/MIX |
| ER total difficulties | 2.07 (0.58) | 1.99 (0.56) | 2.09 (0.58) | 2.65 (0.74) | $F(3, 669) = 38.91$ | $< .001$ | $\eta_p^2 = .149$ | RISK > MIN/BEN/MIX |
| Self-esteem | 5.09 (1.24) | 5.64 (1.20) | 5.69 (1.18) | 5.04 (1.36) | $F(3, 669) = 12.68$ | $< .001$ | $\eta_p^2 = .054$ | BEN/MIX > MIN/RISK |
| Psychological flourishing | 5.09 (1.01) | 5.56 (0.97) | 5.65 (0.84) | 4.99 (1.06) | $F(3, 669) = 20.07$ | $< .001$ | $\eta_p^2 = .083$ | BEN/MIX > MIN/RISK |
| Depression symptoms | 6.03 (4.37) | 4.50 (3.87) | 5.21 (3.77) | 8.56 (5.97) | $F(3, 669) = 26.41$ | $< .001$ | $\eta_p^2 = .106$ | RISK > MIN/BEN/MIX; MIN > BEN; others *n*.s. |
| Anxiety symptoms | 5.61 (4.12) | 3.97 (3.19) | 4.45 (3.33) | 7.33 (4.76) | $F(3, 669) = 26.56$ | $< .001$ | $\eta_p^2 = .106$ | RISK > MIN/BEN/MIX; MIN > |

BEN; others *n*.s.

*Note.* Values are n (%) for categorical variables and M (SD) for continuous variables. MIN = minimal impact; BEN = benefit-driven impact; MIX = mixed impact; RISK = risk-driven impact; HAABI = Human-AI Affective Bonding Inventory; ER = emotion regulation. Continuous variables were tested with one-way ANOVAs and Tukey-adjusted pairwise comparisons. Categorical variables were tested with chi-square or Fisher's exact tests and Holm-adjusted pairwise comparisons. Summarized post hoc results are shown; full pairwise comparisons are available in the supplementary materials.

**Table 5.** *W2 Psychological Functioning Indicators Predicted by W1 Perceived Impact Profiles*

| W2 outcome | MIN $n$ = 34 | BEN $n$ = 68 | MIX $n$ = 120 | RISK $n$ = 51 | Statistic | $p$ | Post hoc |
|---|---|---|---|---|---|---|---|
| Interpersonal needs | 2.49 (0.83) | 2.26 (0.83) | 2.49 (0.98) | 3.27 (1.02) | $F(3, 258) = 0.30$ | .822 | *n*.s. |
| ER total difficulties | 2.05 (0.58) | 2.07 (0.63) | 2.20 (0.73) | 2.75 (0.64) | $F(3, 258) = 0.47$ | .706 | *n*.s. |
| Self-esteem | 4.91 (1.22) | 5.45 (1.10) | 5.42 (1.16) | 4.80 (1.41) | $F(3, 258) = 1.45$ | .227 | *n*.s. |
| Psychological flourishing | 4.97 (0.80) | 5.30 (0.91) | 5.50 (0.80) | 4.86 (0.83) | $F(3, 258) = 2.98$ | .032 | MIX > MIN |
| Depression symptoms | 4.85 (4.18) | 5.45 (3.99) | 6.16 (4.31) | 9.14 (4.61) | $F(3, 258) = 0.59$ | .622 | *n*.s. |
| Anxiety symptoms | 5.41 (3.68) | 5.90 (4.27) | 5.82 (4.08) | 8.00 (4.83) | $F(3, 258) = 0.39$ | .763 | *n*.s. |

*Note.* Values are observed W2 $M$ ($SD$) by W1 profile. Overall profile tests and post hoc comparisons were based on covariate-adjusted models controlling for the corresponding W1 outcome, W2 stressful life events, W1 gender, W1 age, W1 education level, W2 daily AI use time, and W2 AI use frequency. MIN = minimal impact; BEN = benefit-driven impact; MIX = mixed impact; RISK = risk-driven impact; ER = emotion regulation; n.s. = nonsignificant. Observed W2 descriptives were based on W2 completers ($n$ = 273). Adjusted models used complete covariate cases ($n$ = 271), excluding two participants with missing age.

**Figure 1.** *Standardized Means of the Eight Perceived AI-Related Impact Indicators Across the Four-Profile Solution.*

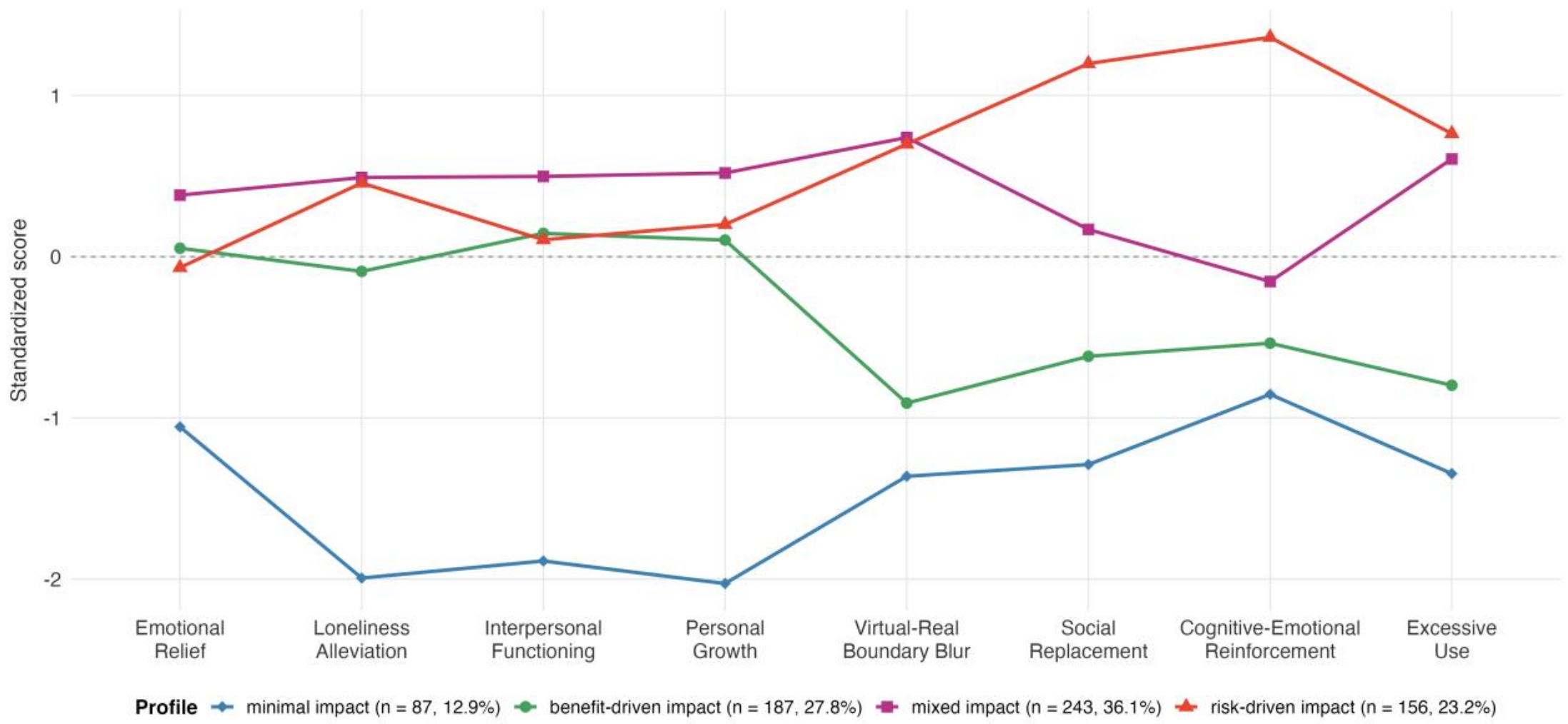

*Note.* Indicators were rescaled to a 0–1 metric and then standardized before model estimation. Higher scores indicate stronger perceived impact.

**Figure 2.** *Human-AI Affective Bonding and Psychological Functioning Across Perceived Impact Profiles*

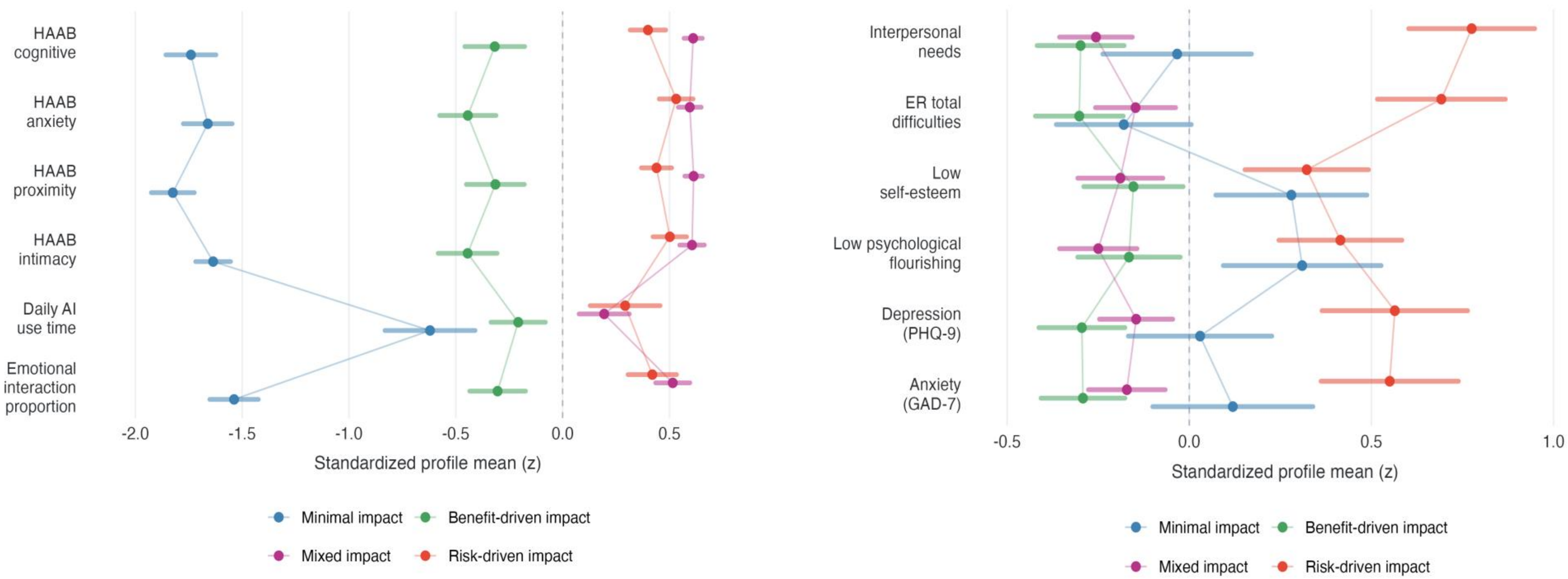

**(a) Human-AI relationship and use characteristics** **(b) Psychological functioning**

*Note.* Points show standardized profile means; intervals show approximate 95% confidence intervals. In Panel B, self-esteem and psychological flourishing were reverse coded so all indicators point toward greater psychological vulnerability. HAABI = Human-AI Affective Bonding Inventory; ER = emotion regulation.

## Appendix A. Semi-Structured Interview Protocol (Study 1)

*Interview Guide for Human-AI Intimate Relationships*

| Section | Topic | Interview Questions and Probes |
|---|---|---|
| **Interaction Experience** | **Relationship Perception and Positioning** | How would you describe the relationship between you and your AI? Please elaborate. |
| | | How do you perceive this relationship? |
| | | If you compare it with human relationships, what type of human relationship does it most closely resemble? In what ways is it different? |
| | **Emotional Experiences and Impressions** | Can you recall the first time you discussed a relatively deep emotional topic with an AI? How did you feel at the time? |
| | | Have you had any particularly memorable interactions, such as experiencing a strong sense of emotional resonance or feeling loved? Could you elaborate? |
| | | Have you ever felt better after talking with an AI? What did the AI do that you found helpful? |
| **Comparison with Human Interactions** | **Human-AI Romantic Relationships** | For participants in a human-AI romantic relationship only: What do you find most appealing about a human-AI romantic relationship? Conversely, what aspects of romantic relationships between humans can never be replicated by AI? |
| | **Significant Others and AI** | Who are the significant others in your offline life? What do you usually talk about with them? In what ways are the manner and content of those conversations similar to or different from your conversations with AI? |
| | | Interviewer note: In the following questions, replace "parents" with the significant other(s) identified by the participant above, where appropriate. |
| | | Are there things you would tell only an AI but would not tell your parents? Are there things you would tell your parents but not an AI? |

| Section | Topic | Interview Questions and Probes |
| --- | --- | --- |
| | | What respective strengths do AI and your parents have in understanding you? In what situations do you feel more comfortable with each of them? |
| | | Do you think your parents and AI could work together to help you regulate your emotions? If so, how should they complement one another? |
| | | Are you currently willing to talk about your emotions with your parents? If not, what prevents you from doing so, and what might AI do to improve the situation? If yes, what do you think prevents some people from discussing their emotions with their parents, and what might AI do to help? |
| **Effects of Human-AI Interaction** | **Self-Perception** | Has emotional interaction with AI changed how you perceive yourself, either positively or negatively? |
| | **Offline Interpersonal Relationships** | How has emotional interaction with AI affected your interpersonal relationships in the offline world? Has it led to any changes in those relationships, either positive or negative? |
| | **Emotional State and Mental Health** | How has emotional interaction with AI affected your emotional state or mental health, either positively or negatively? |
| **Conclusion** | **Future Research** | What suggestions or expectations do you have for future research on human-AI relationships? |
| | **Additional Comments** | Is there anything else you would like to add? |

**Appendix B. Coding Results for Perceived AI-Related Impacts (Study 1)**

*Three-Level Coding Results from the Qualitative Interviews*

| Node ID | Node | Number of Mentions | Subtheme | Main Theme |
|---|---|---|---|---|
| 1 | Improved mood and a more positive emotional state | 23 | **Emotional Relief** | **Positive Perceived Impacts** |
| 2 | Comfort and emotional release after disclosure | 21 | | |
| 3 | Feeling emotionally understood and seen | 8 | | |
| 4 | Emotional stabilization and regaining calm | 7 | | |
| 5 | Feeling loved and emotionally uplifted through romantic interaction | 4 | | |
| 6 | Ongoing companionship and a sense of presence | 12 | **Loneliness Alleviation** | |
| 7 | Relief from loneliness, emptiness, and boredom | 5 | | |
| 8 | Repairing, maintaining, and expanding offline relationships | 17 | **Enhanced Interpersonal Functioning** | |
| 9 | Greater social initiative and social confidence | 11 | | |
| 10 | Greater patience and tolerance toward others and a kinder tone | 10 | | |
| 11 | Learning communication skills and ways of expressing oneself | 9 | | |
| 12 | Greater perspective-taking and understanding of others' viewpoints | 5 | | |

| Node ID | Node | Number of Mentions | Subtheme | Main Theme |
|---|---|---|---|---|
| 13 | More mature ways of interacting with others offline | 2 | | |
| 14 | Helping regulate others' emotions | 1 | | |
| 15 | Greater self-understanding and awareness of emotional needs | 18 | | |
| 16 | Stronger self-affirmation and self-efficacy | 11 | **Personal Growth** | |
| 17 | A more positive mindset | 8 | | |
| 18 | Greater self-acceptance and self-kindness and reduced negative self-perceptions | 7 | | |
| 19 | More authentic self-expression | 6 | | |
| 20 | Personal growth and independence | 5 | | |
| 21 | A more outgoing personality | 5 | | |
| 22 | Broader perspectives and understanding | 2 | | |
| 23 | Greater strength and resilience | 1 | | |
| 24 | Difficulty accepting differing viewpoints | 2 | | **Negative Perceived Impacts** |
| 25 | Excessive self-confidence | 2 | **Cognitive－Emotional Reinforcement** | |
| 26 | Increased desire for control | 1 | | |
| 27 | Amplification of negative emotions | 1 | | |
| 28 | Reduced tolerance for dissent and more rigidly asserted | 1 | | |

| Node ID | Node | Number of Mentions | Subtheme | Main Theme |
|---|---|---|---|---|
| | views | | | |
| 29 | Becoming easily immersed and experiencing withdrawal | 4 | **Excessive Use** | |
| 30 | Becoming absorbed and feeling manipulated | 2 | | |
| 31 | Susceptibility to compulsive or addictive use | 1 | | |
| 32 | Deep integration into everyday life | 1 | | |
| 33 | Losing one's sense of self | 1 | | |
| 34 | Enclosing oneself in a virtual world | 2 | **Virtual－Real Boundary Blur** | |
| 35 | Making real-world commitments with AI | 1 | | |
| 36 | Perceiving AI as a real person and wanting phone contact | 1 | | |
| 37 | Confiding in AI instead of sharing with people offline | 16 | **Social Replacement** | |
| 38 | Higher standards for and comparisons with offline relationships | 5 | | |
| 39 | Reduced need for offline intimacy | 4 | | |
| 40 | Reduced participation in offline social activities | 4 | | |
| 41 | Comparing AI relationships with offline intimate relationships | 2 | | |

| Node ID | Node | Number of Mentions | Subtheme | Main Theme |
|---|---|---|---|---|
| 42 | Greater vigilance and avoidance in offline social interactions | 1 | | |
| 43 | Conflict between AI relationships and offline intimate relationships | 1 | | |
| 44 | Replacing some low-quality social interactions with AI interaction | 1 | | |